\documentclass[twocolumn]{aastex631}

\usepackage[normalem]{ulem}
\usepackage{comment}
\usepackage{amsmath}
\usepackage[utf8]{inputenc}

\newcommand{\pc}{\,\rm{pc}}
\newcommand{\kpc}{\,\rm{kpc}}
\newcommand{\Mpc}{\,\rm{Mpc}}

\newcommand{\Gyr}{\,\rm{Gyr}}
\newcommand{\Myr}{\,\rm{Myr}}

\newcommand{\sig}{$\sigma_{\rm 3D}$}
\newcommand{\siglos}{$\sigma_{\rm los}$}
\newcommand{\sigr}{$\sigma_{\rm r}$}

\newcommand{\gin}{$\gamma_{\rm in}$}
\newcommand{\gout}{$\gamma_{\rm out}$}

\shorttitle{Velocity dispersion profiles of galaxies}
\shortauthors{Han et al.}

\begin{document}

\title{On the Origin of the Variety of Velocity Dispersion Profiles of Galaxies}

\author[0000-0001-9939-713X]{San Han}
\altaffiliation{Email address: sanhan@yonsei.ac.kr}
\affiliation{Department of Astronomy and Yonsei University Observatory, Yonsei University, Seoul 03722, Republic of Korea}

\author[0000-0002-4556-2619]{Sukyoung K. Yi}
\altaffiliation{Email address: yi@yonsei.ac.kr}
\affiliation{Department of Astronomy and Yonsei University Observatory, Yonsei University, Seoul 03722, Republic of Korea}

\author[0000-0002-4731-9604]{Sree Oh}
\affiliation{Department of Astronomy and Yonsei University Observatory, Yonsei University, Seoul 03722, Republic of Korea}
\affiliation{Research School of Astronomy and Astrophysics, Australian National University, Canberra, ACT 2611, Australia}
\affiliation{ARC Centre of Excellence for All Sky Astrophysics in 3 Dimensions (ASTRO 3D), Australia}

\author[0000-0002-5896-0034]{Mina Pak}
\affiliation{School of Mathematical and Physical Sciences, Macquarie University, NSW 2109, Australia}
\affiliation{ARC Centre of Excellence for All Sky Astrophysics in 3 Dimensions (ASTRO 3D), Australia}

\author[0000-0003-2880-9197]{Scott Croom}
\affiliation{ARC Centre of Excellence for All Sky Astrophysics in 3 Dimensions (ASTRO 3D), Australia}
\affiliation{Sydney Institute for Astronomy (SIfA), School of Physics, The University of Sydney, NSW 2006, Australia}

\author[0000-0002-8140-0422]{Julien Devriendt}
\affil{Department of Physics, University of Oxford, Keble Road, Oxford OX1 3RH, UK}

\author[0000-0003-0225-6387]{Yohan Dubois}
\affil{Institut d’Astrophysique de Paris, Sorbonne Université, CNRS, UMR 7095, 98 bis bd Arago, 75014 Paris, France}

\author[0000-0002-3950-3997]{Taysun Kimm}
\affil{Department of Astronomy and Yonsei University Observatory, Yonsei University, Seoul 03722, Republic of Korea}

\author[0000-0001-6180-0245]{Katarina Kraljic}
\affil{Universit\'e de Strasbourg, CNRS, Observatoire astronomique de Strasbourg, UMR 7550, F-67000 Strasbourg, France}

\author[0000-0003-0695-6735]{Christophe Pichon}
\affiliation{Institut d’Astrophysique de Paris, Sorbonne Université, CNRS, UMR 7095, 98 bis bd Arago, 75014 Paris, France}
\affiliation{IPHT, DRF-INP, UMR 3680, CEA, Orme des Merisiers Bat 774, 91191 Gif-sur-Yvette, France}
\affiliation{Korea Institute of Advanced Studies (KIAS) 85 Hoegiro, Dongdaemun-gu, Seoul, 02455, Republic of Korea}

\author[0000-0002-3216-1322]{Marta Volonteri}
\affiliation{Institut d’Astrophysique de Paris, Sorbonne Université, CNRS, UMR 7095, 98 bis bd Arago, 75014 Paris, France}

\begin{abstract}
Observed and simulated galaxies exhibit a significant variation in their velocity dispersion profiles. 
We examine the inner and outer slopes of stellar velocity dispersion profiles using integral field spectroscopy data from two surveys, SAMI (for $z < 0.115$) and CALIFA (for $z < 0.03$), comparing them with results from two cosmological hydrodynamic simulations: Horizon-AGN (for $z = 0.017$) and NewHorizon (for $z\lesssim1$).
The simulated galaxies closely reproduce the variety of velocity dispersion slopes and stellar mass dependence of both inner and outer radii ($0.5\,r_{50}$ and $3\,r_{50}$) as observed, where $r_{50}$ stands for half-light radius.
The inner slopes are mainly influenced by the relative radial distribution of the young and old stars formed in-situ: a younger center shows a flatter inner profile.
The presence of accreted (ex-situ) stars has two effects on the velocity dispersion profiles. 
First, because they are more dispersed in spatial and velocity distributions compared to in-situ formed stars, it increases the outer slope of the velocity dispersion profile.
It also causes the velocity anisotropy to be more radial. 
More massive galaxies have a higher fraction of stars formed ex-situ and hence show a higher slope in outer velocity dispersion profile and a higher degree of radial anisotropy. 
The diversity in the outer velocity dispersion profiles reflects the diverse assembly histories among galaxies.
\end{abstract}

%%%%%%%%%%%%%%%%%%%%%%%%%%%%%%%%%%%%%%%%%%%%%%%%%%%%%%%%%%%%%%%%%%%%%
\section{Introduction}\label{sec:intro}
%%%%%%%%%%%%%%%%%%%%%%%%%%%%%%%%%%%%%%%%%%%%%%%%%%%%%%%%%%%%%%%%%%%%%
Most galaxies in the Universe are expected to exist inside a dark matter halo that extends far beyond the size of the galaxy \citep[e.g.,][]{1980ApJ...238..471R, 1981AJ.....86.1825B}.
In the standard cold dark matter (CDM) paradigm,
dark matter halos are assumed to be affected by various gravitational processes.
Therefore, their structural properties are expected to reflect their assembly and evolutionary history \citep{1996ApJ...462..563N, 2002ApJ...568...52W}.
For example, brightest cluster galaxies (BCGs) and brightest group galaxies (BGGs) located at the center of a large system are surrounded by exceptionally massive and extended dark matter halos \citep{2010ApJ...710..903M, 2018MNRAS.475..648P} presumably as a result of numerous mergers and accretion events during their formation \citep{2004ApJ...614...17G, 2007MNRAS.375....2D}.
However, satellites within such systems are prone to tidal stripping which preferentially removes loosely-bound outer parts of the dark halos, resulting in truncated dark matter halos \citep{2016ApJ...833..109S}.

It has been also claimed that the inner dark matter distributions of dwarf galaxies show flat profiles at the center, deviating from what is expected from cold dark matter properties \citep{1994ApJ...427L...1F, 2005ApJ...621..757S, 2015AJ....149..180O}.
These dark matter profiles with flat centers have revealed the necessity for additional physical processes beyond the simple CDM-based modeling of the Universe.
The effect of baryonic feedback has been proposed as a possible solution. When the feedback is sufficiently strong, it can generate fluctuations in the gravitational potential of the halo, leading to the flattening of the inner dark matter profiles \citep[e.g.,][]{2008A&A...479..123P, 2012MNRAS.421.3464P, 2013MNRAS.429.3068T}.
An alternative scenario that has been suggested involves the self-interaction of dark matter. In this scenario, the collision of dark matter particles within the halo leads to the heating of the central cusp, transforming it into a shallower profile \citep{2000PhRvL..84.3760S, 2013MNRAS.430...81R, 2018PhR...730....1T}.

Therefore, inferring the dark matter profiles of galaxies is useful to understand the evolution of dark halos and their relationship with galaxies, as well as for probing the physical properties of dark matter.
These processes serve as essential tests for the standard cosmological models.

Recent developments in the integral field unit (IFU) technique have enabled the detailed study of the kinematics of galaxies in large quantities and great detail \citep{2012A&A...538A...8S, 2014ApJ...795..158M, 2015MNRAS.447.2857B}.
Specifically, \citet{2022ApJ...930..153S} used the Sydney-AAO Multi-object Integral-field unit system \citep[SAMI,][]{2012MNRAS.421..872C,2015MNRAS.447.2857B,2021MNRAS.505..991C} data to derive dark matter fractions in passive galaxies.
Recent studies suggest that galaxies display a variety of velocity dispersion profiles \citep{2017A&A...604A..30N, 2017A&A...597A..48F, 2018A&A...618A..94P, 2018MNRAS.477..335L, 2018MNRAS.473.5446V, 2019MNRAS.489.3797M, 2020MNRAS.495.4820L, 2020MNRAS.491.2617E}.
The velocity dispersion profile is found to be dependent on the morphology and bulge-to-total ratio of the galaxy \citep{2017A&A...604A..30N} and tends to vary more significantly at larger radii \citep{2018A&A...618A..94P}.
However, the degeneracy between the orbital velocity anisotropy and the total mass profile makes it difficult to reconstruct the dark matter distribution \citep{1982MNRAS.200..361B}.
It is known that this issue can be overcome by using the high-order Gauss-Hermite moment of the line-of-sight velocity distribution $h_4$ \citep{1993MNRAS.265..213G, 2017MNRAS.471.4541R} that is coupled with the velocity anisotropy of the system.
However, there have been reports of certain massive early-type galaxies exhibiting rising velocity dispersion profiles \citep{2020MNRAS.491.2617E} and positive values of $h_4$ \citep{2008MNRAS.390...93K, 2017MNRAS.464..356V, 2018MNRAS.477..335L, 2020MNRAS.496.1857L}.
This appears to be contradictory because rising profiles are usually associated with tangentially-biased anisotropy, whereas positive values of $h_{4}$ are considered hints of radially biased anisotropy.
As a possible solution to this problem, \cite{2018MNRAS.473.5446V} proposed a variation in the total mass profile, and \cite{2020MNRAS.496.1857L} suggested the contribution of intracluster light.

The degeneracy is mainly caused by projection effects, which makes it difficult to interpret the observed velocity dispersion profiles of the galaxies correctly.
Numerical simulations are useful for investigating this issue because they provide detailed kinematic and spatial information on all the particles.
Cosmological simulations provide tests for the $\Lambda$CDM cosmology based on the initial conditions of primordial density fluctuations that are observable in cosmic microwave background data.
Advances in computational resources and simulation codes have enabled the emergence of large, high-resolution cosmological simulations with sophisticated subgrid prescriptions \citep[e.g.,][]{2019MNRAS.483.3336T, 2019MNRAS.490.3196P, 2021A&A...651A.109D}.
Recent numerical simulations have successfully reproduced various observed trends in velocity dispersion profiles \citep[e.g.,][]{2020A&A...641A..60P, 2020MNRAS.495.4820L, 2022MNRAS.513.6134W, 2023MNRAS.520.5651C}.
However, the full understanding of the physical processes that shape galaxy velocity dispersion remains limited.
Variations in the kinematic properties of galaxies are believed to result from the interplay between various processes in the hierarchical assembly paradigm.
Understanding how these processes are encoded in the kinematic properties is crucial.
Conversely, velocity profiles can be used to reconstruct the past assembly and evolution histories of galaxies.

Stars in galaxies typically assemble in two ways: in-situ and ex-situ star formation.
Two types of stellar components exhibit distinct kinematic characteristics.
In-situ formed stars, or simply in-situ stars, exhibit low-velocity dispersion as they retain dynamically cold properties inherited from the viscous nature of the cold gas component from which they originate.
Ex-situ formed stars, or ex-situ stars, are defined as stars that originate from external galaxies and become incorporated into the main galaxy through mergers or accretion events.
During the coalescence process, ex-situ stars lose information about the dynamical properties they had in their original galaxy.
Understanding the distinction between these two components is crucial for comprehending how the velocity dispersion profiles of galaxies are shaped.

This study adopts two cosmological hydrodynamic simulations to measure the velocity dispersion profiles of galaxies using a methodology similar to that of IFU surveys.
By tracing the history of star particles, we aimed to determine the main processes that shaped the current forms of velocity dispersion profiles.
In Section~\ref{sec:data}, we describe the sample selection process used for simulations and observations.
In Section~\ref{sec:slope}, we compare the slopes of the velocity dispersion profiles obtained from simulations and observations.
In Section~\ref{sec:analysis}, we use the Jeans' equation to perform a kinematic analysis and determine the key structural parameters that influence the velocity dispersion profiles of galaxies.
In Section~\ref{sec:discuss}, we discuss the primary factors and processes that determine the velocity dispersion profiles of galaxies.
This study assumes the standard $\Lambda$CDM cosmology based on \cite{2011ApJS..192...18K} (h=0.704, $\Omega_m$=0.272, $\Omega_\Lambda$=0.728) and the stellar initial mass function of \cite{2003PASP..115..763C}.
%%%%%%%%%%%%%%%%%%%%%%%%%%%%%%%%%%%%%%%%%%%%%%%%%%%%%%%%%%%%%%%%%%%%%
\section{Data}\label{sec:data}
%%%%%%%%%%%%%%%%%%%%%%%%%%%%%%%%%%%%%%%%%%%%%%%%%%%%%%%%%%%%%%%%%%%%%

\subsection{Simulation data}\label{sec:simulation}
%%%%%%%%%%%%%%%%%%%%%%%%%%%%%%%%%%%%%%%%%%%%%%%%%%%%%%%%%%%%%%%%%%%%%
We use two cosmological hydrodynamic simulations: Horizon-AGN \citep{2014MNRAS.444.1453D} and NewHorizon \citep{2021A&A...651A.109D}.
Both are based on the same WMAP cosmology (h=0.704, $\Omega_m$=0.272, $\Omega_\Lambda$=0.728) and the hydrodynamic code RAMSES \citep{2002A&A...385..337T}.

Horizon-AGN (hereafter HAGN) involves a periodic cube with a comoving volume of $(142{\Mpc})^3$ and has a ``best'' spatial resolution of $\sim1{\kpc}$.
The vast scale of the simulation allows access to numerous galaxy samples from various environments.
However, the limited gravitational and hydrodynamic spatial resolutions of HAGN, which are comparable to the typical value of the effective radii of dwarf galaxies, result in poor reproduction of sub-kpc scale structures such as thin disks.
Additionally, the mass resolution of $\sim3.5\times10^{6}M_{\sun}$ limits the accurate representation of galaxy kinematics, leading to significant statistical noise in the analysis, particularly when measuring spatially resolved kinematics.
Because these resolution issues become more severe for smaller galaxies, we limit the stellar mass range of our sample to $\log(M_*/M_{\sun}) > 10.75$.
This corresponds to more than 15,000 star particles in each galaxy.
We aim to obtain a statistically reliable number of star particles for measuring the velocity dispersion in 100 equally numbered radial (shell) bins around a galaxy, which corresponds to $\geq 150$ particles in each bin. 
As a result, we managed to include 5,829 galaxies from the HAGN simulation at the final snapshot of $z=0.017$, which includes 40 galaxies with $\log(M_*/M_{\sun}) > 12$ and 405 galaxies with $\log(M_*/M_{\sun}) > 11.5$.

NewHorizon (hereafter NH), which has the best spatial resolution of $\sim 34\pc$, consists of a spherical zoom-in region with a comoving diameter of 20{\Mpc}.
%The minimum redshift of NH presented by \cite{2021A&A...651A.109D} is $z=0.25$.
%We utilize snapshots with lower redshifts, extending down to $z = 0.17$, which have been generated as part of the continuation of the simulation.
We utilize all the snapshots down to $z=0.017$, while the introductory paper by Dubois et al. (2021) for NH used only down to $z=0.25$. 
Because of the use of a zoom-in technique in NH, there is a possibility that some galaxies, usually near the boundary regions of the sphere, contain dark matter particles from low-resolution regions in the initial conditions.
This can increase the shot noise in density, leading to a negative effect on the gravitational stability of the system.
To minimize the contamination effect, we use galaxies without low-resolution dark matter particles for our sample.
To include as many galaxies as possible in our analysis, we extract the galaxies from multiple snapshots in the NH simulation by selecting time intervals of approximately 0.5{\Gyr}.
In total, we selected 12 snapshots in the redshift range of $0.17 < z < 0.96$.
We used a mass cut of $\log(M_*/M_{\sun}) > 9.5$ for the NH sample for easy comparison with the observed data that is described in Section~\ref{sec:data_obs}.
This means that, with the typical masses of star particles of $\sim10^4M_{\sun}$, each galaxy can be resolved into a minimum of 300,000 star particles.
The sample includes 2,104 galaxies in total, with 278 galaxies having $\log(M_*/M_{\sun}) > 10$, and 86 galaxies having $\log(M_*/M_{\sun}) > 10.5$.
While most of our galaxies reside in the field environments, 126 galaxies are located in two group-size halos of the masses of $\log(M_{\rm vir}/M_{\sun}) = 12.83$ and $\log(M_{\rm vir}/M_{\sun}) = 12.96$.
In both simulations, we use the AdaptaHOP algorithm \citep{2004MNRAS.352..376A} to detect galaxies based on the density distribution of star particles.
The evolutionary track of each galaxy is traced by following the main progenitor branch of the merger tree.

NH and HAGN utilize the adaptive mesh refinement (AMR) technique implemented in the RAMSES code, where the hydro and gravitational solvers operate on octree grids.
With each eightfold increase in density, the grid is subdivided into smaller grids.
This approach enables RAMSES to assign smaller grids in higher-density regions to conduct more precise calculations.
In NH, the supplementary refinement criterion is triggered when the Jeans length of gas becomes smaller than four times the size of the cell.
The simulation stops refining grids when it reaches the smallest grid size, i.e., the best spatial resolution.

All stars in simulated galaxies are classified as either ``ex-situ'' or ``in-situ'' stars.
Ex-situ stars are defined as stars that were older than 250{\Myr} at the time of accretion.
The accretion is defined as the moment they were recognized as members of the hosting galaxy by the galaxy finder for the first time.
The utilization of a time threshold helps exclude transient stellar systems within the galaxy, which may sometimes be identified as separate systems through the galaxy finder algorithm.
The choice of a 250{\Myr} time threshold is comparable to the orbital period of the galaxy, ensuring sufficient time for the dissolution of clumps through differential rotation. 
The ex-situ fraction includes both minor and major mergers. Of the two, major mergers are considered to have a more dramatic effect on the velocity dispersion and even morphology of the remnant galaxy \cite[e.g.,][]{2005A&A...437...69B, 2012MNRAS.425.3119H}.
However, we do not distinguish them when identifying ex-situ stars because they have largely the same tendency to increase velocity dispersion, a key property of our investigation.

\begin{figure*}[t]
\begin{center}
\includegraphics[width=0.9\textwidth]{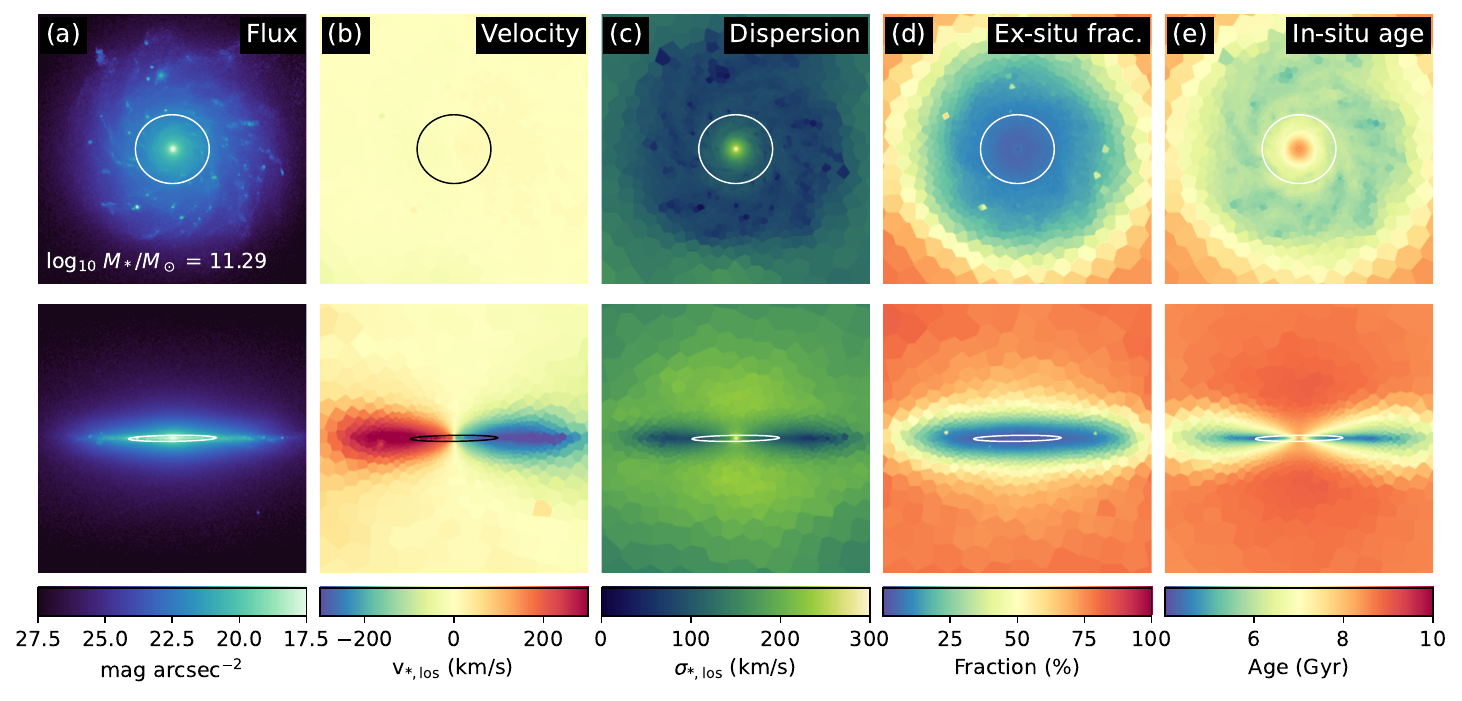}\caption{Mock IFU image of a galaxy in NewHorizon simulation. The face-on (upper row) and edge-on (lower row) projections are shown. Each column represents the projected distribution of r-band flux (a), line-of-sight mean velocity (b) and velocity dispersion (c), ex-situ fraction (d), age of in-situ stars (e). The side length of one panel is 40{\kpc}. The white and black lines indicate the effective ellipse measured in isophote fitting. The difference in velocity dispersion between the old bulge and the young disk leads to a steep increase in velocity dispersion in the inner region.
\label{fig:ifu_sample_large}}
\end{center}
\end{figure*}

\begin{figure*}[t]
\begin{center}
\includegraphics[width=0.9\textwidth]{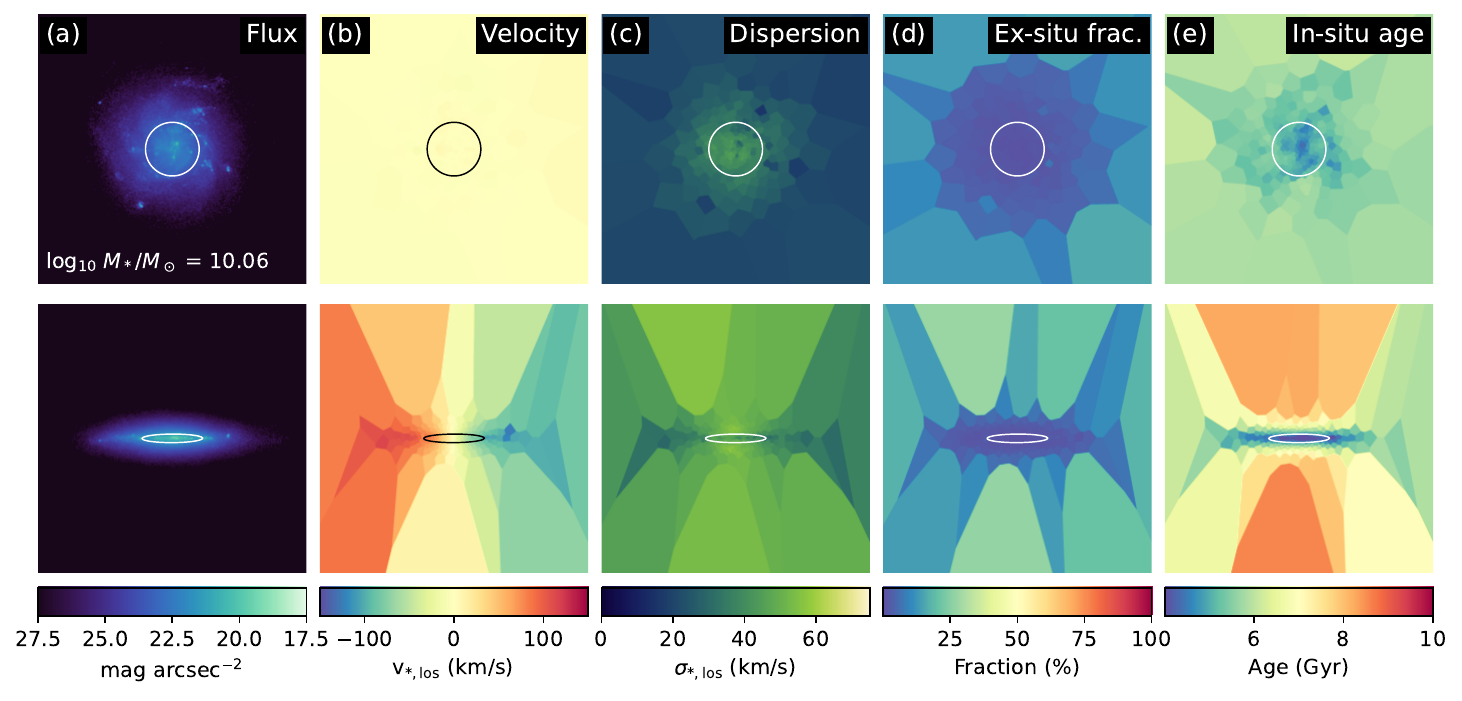}\caption{Mock IFU images simiar to Figure~\ref{fig:ifu_sample_large}, with a less massive galaxy from NewHorizon. The side length of one panel is 40{\kpc}. 
Compared to its more massive counterpart in Figure~\ref{fig:ifu_sample_large}, the galaxy contains younger stars in the inner region, resulting in a relatively flat velocity dispersion profile at the center.
\label{fig:ifu_sample_small}}
\end{center}
\end{figure*}

Finally, a mock line-of-sight velocity dispersion (\siglos) profile is generated for each galaxy using stars grouped according to Voronoi-tessellation regions (See details in Section~\ref{sec:postproc}).
Figures~\ref{fig:ifu_sample_large} and \ref{fig:ifu_sample_small} show randomly selected high and low-mass disk galaxies in NH.
The two-dimensional (2D) images of stellar SDSS $r$-band flux (panel (a)), line-of-sight mean velocity (panel (b)), velocity dispersion (panel (c)), fraction of stars with ex-situ origin (panel (d)) and mean age distribution of stars formed in-situ (panel (e)) are shown in two different projections (face-on and edge-on).

\subsection{Observation data}\label{sec:data_obs}
This section describes the observational data collected from the SAMI and Calar Alto Legacy Integral Field Area (CALIFA) surveys that are used to compare gradients of the stellar velocity dispersion.
We employed the third release data from the SAMI survey \citep{2021MNRAS.505..991C}.
The SAMI \citep{2012MNRAS.421..872C,2015MNRAS.447.2857B} employs 13 fused optical fiber bundles (hexabundle), each containing 61 1.6\arcsec diameter fibers \citep{2011OExpr..19.2649B,2014MNRAS.438..869B}, feeding the AAOmega dual-arm spectrograph mounted on the Anglo-Australian Telescope \citep{2006SPIE.6269E..0GS}. The blue and red arms use 580V and 1000R gratings, covering, 3750--5750\AA\ at a resolution of R=1808, and 6300--7400\AA\ at a resolution of R=4304, respectively. 
The SAMI survey includes more than 3,000 galaxies with a stellar mass range of $\log(M_*/M_\sun) = 8$--12 and a redshift range of $0.004 < z < 0.115$ \citep{2015MNRAS.447.2857B,2021MNRAS.505..991C}. More than two-thirds of the SAMI samples are from three equatorial fields (G09, G12 and G15) of the Galaxy And Mass Assembly survey \citep{2011MNRAS.413..971D}. In addition, galaxies from eight clusters have been observed to complete the environmental matrix \citep{2017MNRAS.468.1824O}. 
We use spatially-resolved line-of-sight stellar velocity dispersions published by the SAMI team \citep{2021MNRAS.505..991C}. \cite{2017ApJ...835..104V} described the measurement of SAMI stellar kinematics using Penalised piXel-Fitting software \citep[pPXP;][]{2004PASP..116..138C,2017MNRAS.466..798C}. We select 2,146 galaxies with stellar mass $\log(M_*/M_\sun) > 9.5$ to reliably estimate the velocity dispersion gradient.
We use the stellar masses provided by the SAMI catalog \citep{2015MNRAS.447.2857B}, derived from i-band magnitudes and g-i colors.

Partial data in our sample is derived from the CALIFA survey \citep{2012A&A...538A...8S,2013A&A...549A..87H} as well.
Observations were made in the PMAS/PPAK spectrograph \citep{2005PASP..117..620R,2006PASP..118..129K} using a 3.5-m telescope at the Calar Alto observatory. The field of view of the PPAK is 74'' × 64'', which comprises 382 fibers of a 2.''7 diameter each \citep{2006PASP..118..129K}.
The galaxies were observed with two spectroscopic setups, using the gratings V500 with a resolution ($\lambda/\Delta\lambda$) of R$\sim$850 at 5000\AA\ (FWHM$\sim$6\AA) covering 3745--7500\AA, and V1200 with a resolution of R$\sim$1650 at 4500\AA\ (FWHM$\sim$2.7\AA), covering 3650--4840\AA.
In this study, we selected 501 galaxies, observed with the V500 setup, with stellar masses $\log(M_*/M_\sun)$ from 9.5 to 11.5 and redshifts $0.005 < z < 0.03$.
The stellar masses of CALIFA galaxies are derived from the NASA-Sloan Atlas \citep{2011AJ....142...31B}.

In addition to the SAMI and CALIFA survey data, we use published data from two other sources.
\cite{2018MNRAS.473.5446V} examined 90 early-type galaxies at the fixed radii of 2{\kpc} and 20{\kpc} using the MASSIVE IFU survey. 
We also use the line-of-sight velocity dispersion ({\siglos}) slopes and K-band magnitudes derived through long-slit spectroscopy of BCGs and BGGs from \citet[Figure 5]{2018MNRAS.477..335L}.
For these catalogs, we employ the stellar mass estimated from K-band magnitude using the relation provided by \cite{2013ApJ...778L...2C}.

We note that in our main results, we did not apply any morphological classifications on the simulated galaxies.
For observations, SAMI and CALIFA encompass all morphological types of galaxies, whereas other sources, such as \cite{2018MNRAS.473.5446V} and \cite{2018MNRAS.477..335L}, include only early-type galaxies.
This bias is somewhat alleviated at the high-mass end, where both the SAMI and CALIFA samples are dominated by early-type galaxies.
Although galaxy morphology has not been directly employed in our main results, discussions on the dependence on morphology are included in Appendix \ref{sec:morph}.
Table \ref{tab:sample} summarizes the overall properties of simulation and observation data, presenting the total number of galaxies, redshift range, stellar mass range, spatial resolution, and type of galaxies.

\begin{deluxetable*}{lrrrrc}
\tabletypesize{\footnotesize}
\tablewidth{0pt}
\tablecaption{ Summary of sample data \label{tab:sample}}
\tablehead{
\colhead{Name} & \colhead{\# of galaxies} & \colhead{$z$} & \colhead{$\log (M_*/M_{\odot})$} & \colhead{dx\tablenotemark{$\dagger$} (kpc)} & \colhead{Galaxy type}
}
\decimals
\startdata
NH\tablenotemark{s} & 2,104 &  0.17--0.96 & 9.5--11.4 & 0.034 & All \\
HAGN\tablenotemark{s} & 5,829 &  0.017 & 10.75--12.7 & 1 & All \\
SAMI\tablenotemark{o} & 2,146 &  0.004--0.115 & 9.5--11.9 & 1 & All \\
CALIFA\tablenotemark{o} & 501 &  0.005--0.03 & 9.5-11.5 & 0.38 & All \\
MASSIVE\tablenotemark{o} & 90 &  $<0.025$ & 11.6--12.2 & 0.8 & ETGs \\
\cite{2018MNRAS.477..335L}\tablenotemark{o} &  70  & 0.05--0.3 & 11.1--12.7 & - & BCGs (BGGs) \\
\enddata
\tablenotetext{s}{Simulation data}
\tablenotetext{o}{Observation data}
\tablenotetext{\dagger}{Corresponds to the best spatial resolution for the case of simulations and rough estimates of HWHM at the mean distance for the case of observations.}
\tablecomments{For simulated 2-D datasets, sample sizes are multiplied by 24 after employing different line-of-sight projections.}
\vspace{-0.5cm}
\end{deluxetable*}

%%%%%%%%%%%%%%%%%%%%%%%%%%%%%%%%%%%%%%%%%%%%%%%%%%%%%%%%%%%%%%%%%%%%%
\section{Slope of \texorpdfstring{$\sigma$}{σ} profile}\label{sec:slope}
%%%%%%%%%%%%%%%%%%%%%%%%%%%%%%%%%%%%%%%%%%%%%%%%%%%%%%%%%%%%%%%%%%%%%
\subsection{Post-processing simulation galaxies}\label{sec:postproc}
This section describes the post-processing procedures conducted on the simulation galaxies for producing 2D velocity dispersion maps for comparison with the IFU-observed data.
In the simulations, a spherical boundary is defined for each galaxy with a radius where the projected surface brightness drops below $26.5\,\mathrm{mag}/\mathrm{arcsec}^2$.
This is analogous to the conventional cut for the boundary of galaxies in photometry.
The luminosity of each star particle is evaluated based on the population synthesis model of \cite{2003MNRAS.344.1000B}, in combination with the stellar initial mass function of \cite{2003PASP..115..763C}.
The radius is measured as the average of three different line-of-sight directions (x, y, and z).

For each galaxy, we measure the stellar line-of-sight velocity dispersion projected onto a 2D plane.
First, 24 random directions were selected for each galaxy.
In each projected view, the star particles are binned into segments of Voronoi tessellation.
We design the Voronoi cells to contain an approximately equal number of star particles, following \cite{2003MNRAS.342..345C}.
The effective number of stars per cell is 5,000 for NH and 100 for HAGN.
The former and latter correspond to the stellar masses of the $\sim5\times10^{7}M_{\sun}$ and $\sim3.5\times10^{8}M_{\sun}$, respectively, which are comparable to the typical sizes of the SAMI spaxels.
For each cell, the $r$-band flux-weighted standard deviation of the stellar line-of-sight velocities is measured.
It is worth noting that the velocity dispersion we examine is consistent with that measured in IFU observations, which differs from the traditional term of that derived from aperture spectroscopy.
The former measures the standard deviation of stellar velocity in a localized region, while the latter measures the integrated second velocity moment across the entire galaxy, considering spatial gradients in the mean velocity as well.

\subsection{Measuring the \texorpdfstring{$\sigma$}{σ} slope}
\label{sec:sigma_slope}
%%%%%%%%%%%%%%%%%%%%%%%%%%%%%%%%%%%%%%%%%%%%%%%%%%%%%%%%%%%%%%%%%%%%%

To measure the radial velocity dispersion profile for each projected view of observed and simulated galaxies, an effective ellipse is computed to enclose half of the total flux of the galaxy.
The Kinemetry method \citep{2006MNRAS.366..787K} is employed to determine the effective ellipse that follows the isophotal lines.
The log r-band flux is assumed to be even moments (n = 0, 2, 4) of the harmonic function.
While maintaining the orientation and ellipticity of the effective ellipse, from 20 to 50 concentric ellipse bins with different sizes are selected.
The radial velocity dispersion profiles are then measured as a function of the semi-major axes of the ellipse bins.

For the systematic measurement of the velocity dispersion profile, we adopt the broken power-law fit of \cite{2018MNRAS.473.5446V},
\begin{equation}\label{eq:bpowerlaw}
\sigma_{f}(r)=\sigma_0 2^{\gamma_1-\gamma_2}\left(\frac{r}{r_{b}}\right)^{\gamma_1}\left(1+\frac{r}{r_{b}}\right)^{\gamma_2-\gamma_1},
\end{equation}
where $r$ represents the semi-major axis of the ellipse bin, $\gamma_1$ and $\gamma_2$ represent the inner and outer asymptotic power-law gradients of the fitted curve, respectively, and $r_{b}$ is the break radius, and $\sigma_0$ is the normalization of the profile (value of $\sigma_{f}$ at $r=r_{b}$).
The fit is performed using \texttt{optimize.curve\_fit} function in \texttt{scipy} module, by minimizing the chi-square in the log-log plane using four parameters, $\gamma_1$, $\gamma_2$, $r_{b}$, and $\sigma_{0}$.
To reduce potential issues arising from beam-smearing in observations and spatial resolution in simulations, we exclude the data points within a radius of $0.1r_{50}$, where $r_{50}$ denotes the semi-major axis of the effective ellipse\footnote{The median values of $r_{50}$ are roughly 2{\kpc} and 6{\kpc} for NH and HAGN, respectively, while the gravitational force calculation resolutions are 34{\pc} and 1{\kpc}.} 

The slope of the fitting function in the log-log plane,
\begin{equation}\label{eq:gamma}
\gamma(r) = \frac{d(\ln{\sigma_{f}})}{d(\ln{r})} = \frac{\gamma_1+\gamma_2(r/r_{b})}{1+r/r_{b}},
\end{equation}
is measured at two radial points.
The inner slope is defined as the slope at $0.5\,r_{50}$, denoted as $\gamma_{\rm in}=\gamma(0.5\,r_{50})$, and the outer slope is defined as the slope at the $3\,r_{50}$, denoted as $\gamma_{\rm out}=\gamma(3\,r_{50})$.

We use inner and outer slopes measured from \cite{2018MNRAS.473.5446V}.
Because they presented slopes only at 2 and 20{\kpc}, while the break radius, $r_b$, was fixed at 5{\kpc}, the broken power-law curve can be recovered by using the following equations,
\begin{equation}\label{eq:convert}
\begin{aligned}
\gamma_1 &= (14 \gamma_{i} - 5 \gamma_{o})/9 \\
\gamma_2 &= (25 \gamma_{o} - 7 \gamma_{i})/18,
\end{aligned}
\end{equation}
where $\gamma_i$ and $\gamma_o$ denote slopes measured at 2 and 20{\kpc}, respectively.
After recovering $\gamma_1$ and $\gamma_2$, we re-measured slopes at $0.5\,r_{50}$ and $3\,r_{50}$ to ensure consistency with our definitions of the inner and outer radii.
We derived the half-light radius of each galaxy from the published catalog of \cite{2019ApJ...874...66G}.

We also utilize the slopes of velocity dispersion profiles presented in \cite{2018MNRAS.477..335L}, Figure 5.
Given that the slope has been measured from a single power-law fit, we use the same values for both the inner and outer radii in our analysis.

\begin{figure}[t]
\begin{center}
\includegraphics[width=0.47\textwidth]{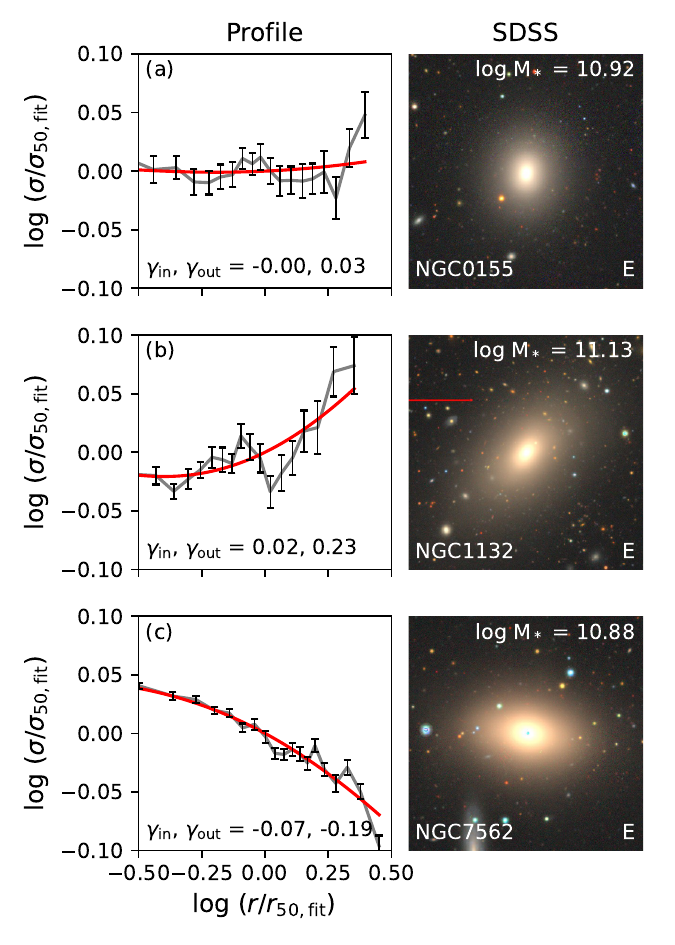}
\caption{
Examples of velocity dispersion profiles of CALIFA galaxies, with their photometric image. The left column shows the radial trend of velocity dispersion (black), with a fitted broken power-law curve (red). The right column shows the SDSS image of target galaxies with stellar mass, name, and morphology. 
Galaxies with similar visual morphology exhibit different trends in their velocity dispersion profiles.
Galaxy morphologies are derived from the same dataset used in \cite{2019ApJ...880..149P}, where visual classification was performed.
\label{fig:sample_profile}}
\end{center}
\end{figure}
Figure~\ref{fig:sample_profile} demonstrates the variation in the {\siglos} profiles among elliptical galaxies.
The three sample galaxies extracted from the CALIFA survey exhibit flat, rising, and falling profiles. 
Even for similar morphologies and masses, the profiles are significantly different.

We select the final sample from the broken power-law fitted data based on the reduced chi-square and mean $\sigma$/error ratio, where ``error'' means the error in {\siglos}.
This secures the reliability of the data and avoids over or under-fitted samples due to poor measurement of the data or complex behaviors of the {\siglos} profile caused by interlopers, mergers, and close companions.
Detailed procedures and criteria are described in Appendix~\ref{sec:fitting_quality}.
The final dataset comprises 2,424 galaxies, with 2,101 from SAMI and 323 from CALIFA.
The fitted simulation data consist of separate line-of-sight views, with 20,985 from NH and 138,896 from HAGN.

\begin{figure}[t]
\begin{center}
\includegraphics[width=0.47\textwidth]{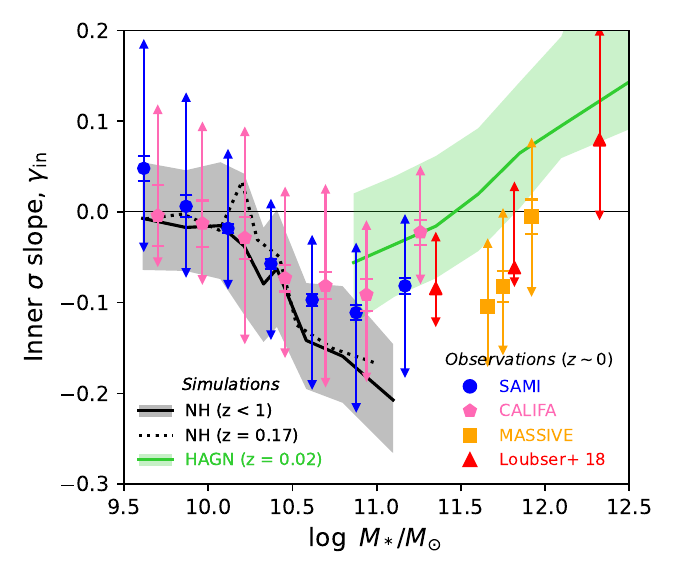}
\caption{
The inner velocity dispersion slope measured at $0.5\,r_{50}$ as a function of galaxies' stellar mass. Solid lines show full sample of our simulated galaxies, with shades indicating the $1\sigma$ scatter. The data for NH galaxies from the final snapshot ($z=0.17$) is shown as a dotted line. Observed data are also presented: $1\sigma$ scatters are shown by arrows and the standard errors of the median are shown by bars.
A fall-and-rise trend can be seen with a turning point at $\log(M_*/M_\sun)\sim11$.
\label{fig:mstar_vs_gamma_all}}
\end{center}
\end{figure}

Figure~\ref{fig:mstar_vs_gamma_all} shows the inner slope, {\gin}, measured in simulations and observations as a function of stellar mass.
The black and green lines represent galaxies from NH and HAGN, respectively.
Observational data are also presented.
The most notable feature of this diagram is the ``V'' shape.
The inner slope of the {\siglos} profile becomes steeper (more negative slope) with increasing stellar mass until $\log(M_*/M_\sun)\sim11$, after which it starts increasing. 
This fall-and-rise trend appears in both the simulated and observed samples.
The most representative samples in this diagram are the SAMI and CALIFA data because they contain a large number of galaxies.
They both show a ``V'' trend.

The ``V'' trend is visible in the simulations as well.
NH does not show the upturn above $\log(M_*/M_\sun)\sim11$ because the field-environment simulation is almost exclusively composed of late-type galaxies.
If only late-type galaxies are selected from the observation, the trend shows a monotonic decrease with stellar mass down to $\gamma_{\rm in}\sim-0.2$, which is in good agreement with NH. The resulting figure is shown in Figure~\ref{fig:mstar_vs_gamma_ltg}-(b) in Appendix.
HAGN has a large volume containing a cosmologically representative sample of galaxies and exhibits an upturn trend in the high mass range. 
In contrast, the kinematic structure of its low-mass galaxies below our mass cut ($\log(M_*/M_\sun)<10.75$) may not be suitable for this analysis, as we explained in Section~\ref{sec:simulation}; therefore, we excluded them from the diagram.
If we accept the low-mass galaxies from the high-resolution NH simulation and massive galaxies from the HAGN simulation as a combination, we can reproduce the observed fall-and-rise trend.
The origins of this trend are discussed in the following section.

One may find the offset of the observed data from MASSIVE uncomfortably large. 
The {\em vertical} offset may have originated from a {\em horizontal} offset, i.e., in stellar mass estimates for observed galaxies.
Alternatively, the HAGN simulation may be incorrect by as much as the vertical offset, which is not inconceivable considering its relatively poor (1{\kpc}) resolution. 
On the other hand, we use only the most massive galaxies from HAGN, exactly worrying about the issue, and the offset persists even for the most massive galaxies for which the signals (or calculations) should be more reliable than for less massive galaxies. 

\begin{figure}[t]
\begin{center}
\includegraphics[width=0.47\textwidth]{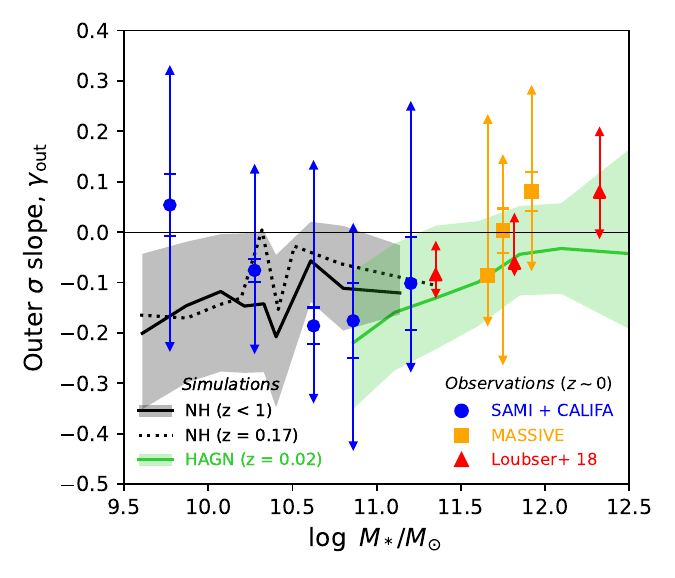}
\caption{The outer slope measured at $3\,r_{50}$ as a function of galaxy's stellar mass.
Similar schemes with Figure~\ref{fig:mstar_vs_gamma_all} are employed, except that the SAMI and CALIFA samples are combined (blue symbols). Overall, the outer slope exhibits a large variation. Massive galaxies ($\log(M_*/M_\sun) \gtrsim 11$) exhibit a hint of an increasing slope with stellar mass.
\label{fig:mstar_vs_gamma_out}}
\end{center}
\end{figure}

Figure~\ref{fig:mstar_vs_gamma_out} shows the outer slope, {\gout}, as a function of stellar mass.
The black and green lines represent the projected outer slopes and the 1$\sigma$ scatters for NH and HAGN galaxies, respectively.
The observed data is also presented.
We show the combined data of SAMI and CALIFA in this figure because SAMI and CALIFA by themselves show extremely large scatters in outer slopes.
The number of galaxies in the drawing sample is reduced to 322 (190 from SAMI and 132 from CALIFA) after selecting observations that consist of data reaching up to $3\,r_{50}$ from the center.
The large scatters may be indicative of a large variation in the physical properties in the outskirts of galaxies.
In the mass range of $\log(M_*/M_\sun)\lesssim11$, NH exhibits no distinct trend.
SAMI and CALIFA observations show a weak decreasing trend, but the errors are so large that the trend may not be statistically relevant.
In addition, the ``trend'' might have originated from the continuation of the broken power-law fits that are dominated (in terms of chi-square evaluation) by the brighter inner profiles.
More massive galaxies ($\log(M_*/M_\sun) \gtrsim 11$)
show a hint of an increasing trend with stellar mass, and it is consistently visible for both observed and simulated galaxies.
We will discuss the origin of this trend in the following sections. 

\subsection{\texorpdfstring{$\sigma$}{σ} profiles and assembly history}\label{sec:sigm_assembly}
In the previous section, we used 2D properties to compare the simulations with observations and demonstrated their similar behavior with respect to stellar mass.
However, projection effects cause complications in analysis and interpretation.
To measure the intrinsic kinematic properties of the galaxies in 3-D space, we set 100 spherical bins (shells) centered on each simulated galaxy with equally numbered star particles.
The luminosity-weighted velocity dispersion in each shell is measured as
\begin{equation}\label{eq:sig_mean}
\sigma_{\rm 3D} = \sqrt{(\sigma_{\rm r}^2+\sigma_\theta^2+\sigma_\phi^2)/3},
\end{equation}
where {\sigr}, $\sigma_\theta$, and $\sigma_\phi$ are the velocity dispersion of stars in three directions in spherical coordinates.

\begin{figure}[t]
\includegraphics[width=0.47\textwidth]{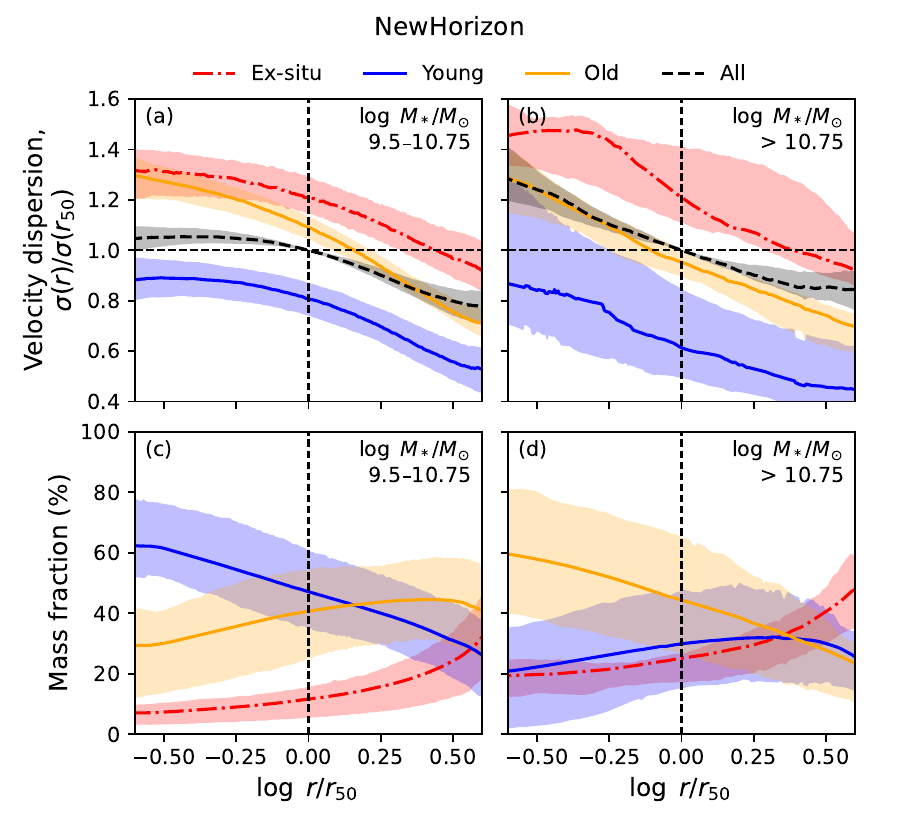}\caption{Radial velocity dispersion (\sig) profiles (panels (a) and (b)) and mass fraction profiles (panels (c) and (d)) of the NH galaxies.
The left and right panels are for two different mass ranges. 
The whole stellar data (black dashed line) is divided into young in-situ (blue solid line), old in-situ (orange solid line), and ex-situ stars (red dash-dotted line).
An age cut of 3{\Gyr} separates the young and old stars.
The shades indicate the $1\sigma$ scatters.
Low-mass and high-mass galaxies exhibit an inverted radial composition of young and old stars.
\label{fig:sig_orig}}
\end{figure}

\begin{figure}[t]
\includegraphics[width=0.47\textwidth]{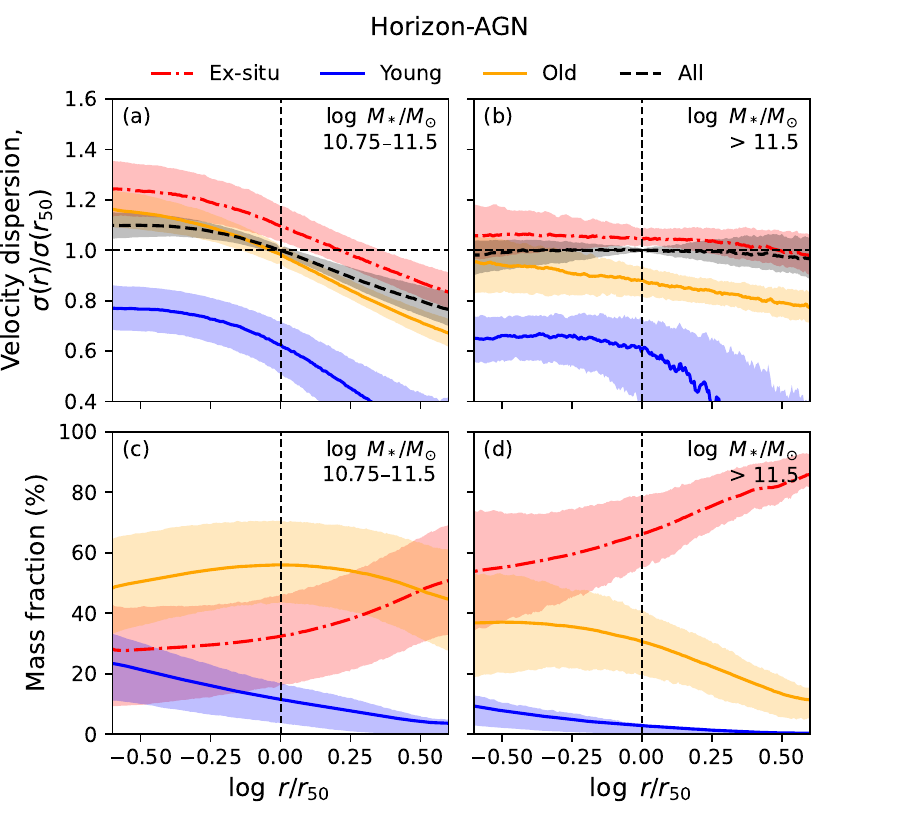}\caption{
Similar to Figure~\ref{fig:sig_orig}, but based on the HAGN sample with different stellar mass ranges, $\log(M_*/M_\sun) = 10.75$--11.5 and $\log(M_*/M_\sun) > 11.5$.
Most massive galaxies in the simulation are dominated by ex-situ stars, and the galaxy's overall velocity dispersion profile is constant.
\label{fig:sig_orig_hagn}}
\end{figure}

To figure out the origin of the different shapes in the velocity dispersion profile, it is important to trace the assembly and formation history of the stellar components of galaxies.
For this purpose, we plot Figures~\ref{fig:sig_orig} and \ref{fig:sig_orig_hagn}, which show the velocity dispersion profiles and mass fractions of stellar components from different origins.
In Figure~\ref{fig:sig_orig}, we plot the {\sig} profiles (panels (a) and (b)) and mass fractions (panels (c) and (d)) of the in-situ and ex-situ stars as a function of the radial distance in the NH galaxies.
The in-situ stars are divided into young and old populations based on an arbitrary age cut of 3{\Gyr}.
The left and right columns represent the two different ranges of stellar masses.
Ex-situ stars generally exhibit higher velocity dispersion at all radii and thus increase the mean velocity dispersion in the region.
Because ex-situ stars are more predominant in the outer regions of the galaxy \citep{, 2016MNRAS.458.2371R}, the strongest impact occurs in the outer regions.
This is consistent with the formation of dynamically hot stellar components via accretion \citep{2006MNRAS.365..747A, 2016MNRAS.463.3948D, 2019ApJ...883...25P, 2022A&A...660A..20Z}.

In contrast, in-situ stars show relatively lower velocity dispersion overall \citep{2016MNRAS.463.3948D}.
When comparing samples of different ages, older stars exhibit higher values of {\sig}.
This result can be interpreted in the context of dynamical heating \citep{1993ApJ...403...74Q,2024ApJS..271....1Y}.
Newly formed stars are likely to retain dissipative gas dynamics with low {\sig}; however, their velocity dispersion is increased with time due to continuous gravitational perturbations from internal or external sources such as spiral arms, giant molecular clouds, mergers, and galaxy encounters.
Consequently, older stars tend to exhibit dynamically hotter kinematics \citep{1977A&A....60..263W, 2016MNRAS.462.1697A, 2021ApJS..254....2P, 2021MNRAS.506.1761S}.

Panels (c) and (d) show that old and young stars have different radial distributions for the two samples with different masses.
In low-mass galaxies (panel (c)), the inner region is increasingly occupied by young stars, reducing {\sig} in the inner region, and the fraction of old stars increases with increasing radial distance, raising {\sig} gradually.
As a result, the {\sig} slope of the inner region is flattened.

In the more massive galaxies (panel (d)), the trend is reversed: the inner region is dominated by old in-situ stars, which increases the central {\sig}, and the fraction of young stars remain relatively stable (or slightly increases) with the radial distance, leading to a steeply falling {\sig} profile.
This may indicate a transition in the order of the star formation process from outside-in to inside-out with increasing stellar mass \citep{2015ApJ...804L..42P, 2019ApJ...872...50L}.
Therefore, the inner {\sig} profiles are determined by the relative radial distributions of the young and old stellar populations.
This can also be seen in mock IFU images.
Figure~\ref{fig:ifu_sample_large} shows a massive galaxy containing an old core with high {\siglos}, surrounded by a young disk with low {\siglos} (panel (e)), exhibiting a steep inner gradient in {\siglos} map (panel (c)).
In contrast, Figure~\ref{fig:ifu_sample_small} shows a lower-mass galaxy containing young in-situ stars in the central region and age increases with radial distance (panel (e)), exhibiting a shallow inner gradient in {\siglos} map (panel (c)).

Figure~\ref{fig:sig_orig_hagn} is the counterpart of Figure~\ref{fig:sig_orig} using HAGN galaxies and different mass ranges.
The relative importance between the three components, as shown in panels (a) and (b), is similar to that of NH, with the ex-situ stellar component having the highest and young in-situ stars having the lowest.
Similar to NH, the fraction of ex-situ stars increases with radial distance in panels (c) and (d).
Flattened {\sig} profiles are observed in more massive galaxies (panel (b)), primarily following the distribution of ex-situ stars that dominate the composition of the galaxy.
For the in-situ components, the relative fraction of old stars over young stars increases with increasing radial distance, which is understandable in the context of dynamical heating and migration.
This is in contrast to the high-mass sample of NH shown in panel (d) of Figure~\ref{fig:sig_orig}, which has a similar stellar mass range.
This discrepancy has two possible explanations. 
First, the morphologies of high-mass end galaxies in NH are dominated by late-type galaxies which contain an old bulge with a star-forming disk. 
In contrast, HAGN includes a substantial number of early-type galaxies in this mass regime, which likely have central star formation rather than extended disk-mode star formation.
Second, HAGN lacks sufficient resolution to accurately distinguish between bulge and disk components, particularly for lower-mass galaxies ($\log(M_*/M_\odot) \lesssim 11$).

\begin{figure}[t]
\includegraphics[width=0.47\textwidth]{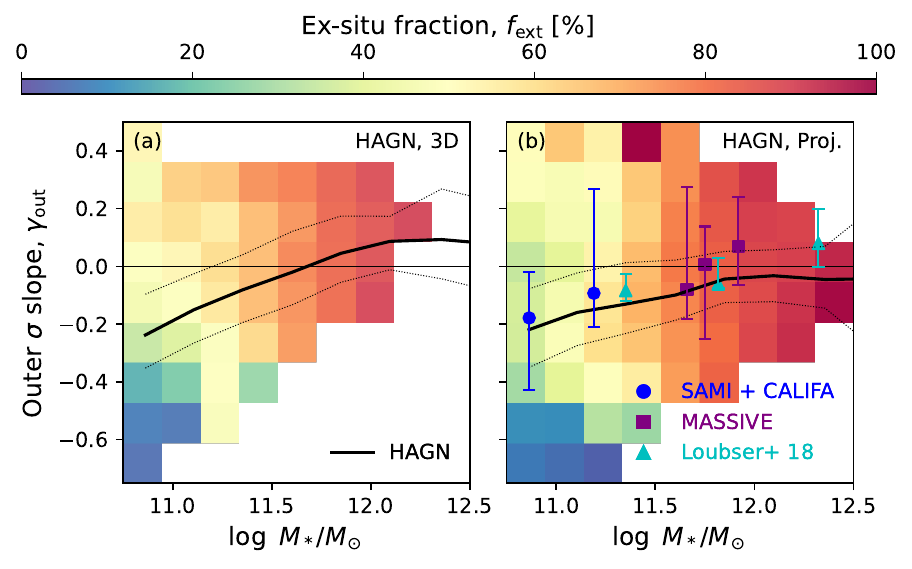}
\caption{Outer slopes of velocity dispersion profile as a function of stellar mass in the range of $\log(M_*/M_\sun)=10.75$--12.25, which corresponds to the massive end of Figure~\ref{fig:mstar_vs_gamma_out}, for the HAGN galaxies.
Outer slopes before (a) and after (b) applying projection effects are shown.
Color pixels in the background represent the median ex-situ fraction.
More massive galaxies exhibit higher ex-situ fractions. Galaxies with higher ex-situ fractions have increased outer velocity dispersion slopes.
\label{fig:facc_slope}}
\end{figure}

To quantify the effect of external accretion, we define ex-situ fraction as the mass fraction of ex-situ stars.
Figure~\ref{fig:facc_slope} shows the impact of ex-situ fraction on the outer slope in the same format as Figure~\ref{fig:mstar_vs_gamma_out}, based on HAGN galaxies. 
Panels (a) and (b) show the slopes of the intrinsic (\sig) and projected (\siglos) profiles, respectively. 
The same broken power-law fit and slope measurement described in Equations \ref{eq:bpowerlaw} and \ref{eq:gamma} are applied to {\sig} by using $r$ as the radial distance from the center.
Panel (b) also presents the observed data. 
As shown in Figure~\ref{fig:mstar_vs_gamma_out}, a mild positive correlation exists between {\gout} and the stellar mass in Panel (b).
The mass dependence is even more evident in the intrinsic 3D case, as shown in Panel (a). 
The mass trend is related to the impact of ex-situ fraction.
In Panel (b), it is apparent that more massive galaxies have higher ex-situ fraction values.
This is expected in the hierarchical paradigm, in which more massive galaxies are likely to have experienced a larger number of mergers and accretions \citep{2013ApJ...766...38L, 2016MNRAS.463.3948D, 2016MNRAS.458.2371R, 2020MNRAS.497...81D, 2022ApJ...935...37R}.
The apparent correlation between {\gout} and ex-situ fraction can also be explained by the fact that ex-situ stars are more spatially dispersed and kinematically hotter than in-situ stars (as shown in Figure~\ref{fig:sig_orig}).
This dominance of ex-situ stars at the outskirts leads to an increase in {\gout} towards zero.
The presence of a vertical gradient, although relatively less pronounced in Panel (b) due to the projection effect, suggests that the variation of ex-situ fraction among galaxies causes the diversity of $\sigma$ profiles of outer halos.

\section{Kinematic analysis of the velocity dispersion profile}\label{sec:analysis}

This section presents the investigations of the structural properties of galaxies that govern the shape of the $\sigma$ profile.
To address this in a quantitative way, we employ the Jeans' equation in spherical coordinates \citep[4.215]{2008gady.book.....B}, 
\begin{equation}\label{eq:binney}
\frac{d(\rho_{*}\overline{v_r^2})}{dr}+2\frac{\beta}{r}\rho_{*}\overline{v_r^2}=-\rho_*\frac{d\Phi}{dr},
\end{equation}
where $\rho_*$ is the stellar density of the system, $\overline{v_r^2}$ is the second moment of velocity in the radial direction, $\Phi$ is the gravitational potential of the system, and $\beta$ is the anisotropy parameter defined as
\begin{equation}\label{eq:anisotropy}
\beta = 1 - \frac{\overline{v_\theta^2} + \overline{v_\phi^2}}{2 \overline{v_r^2}}.
\end{equation}
Here, $\overline{v_r^2}$, $\overline{v_\theta^2}$, and $\overline{v_\phi^2}$ represent the second moment of velocities of stars in spherical coordinates.
It is worth noting that our definition of $\beta$ considers not only the anisotropy of the velocity dispersion but also the bulk tangential velocity of the system (i.e., rotation).
This implies that $\beta$ determines whether the system is supported tangentially or radially.

We applied an assumption that the system has an equilibrium of inflow and outflow in the radial direction, $\overline{v_r} = 0$, so that $\overline{v_r^2} = \sigma_{\rm r}^2$, and the coordinate axis is aligned to the direction of the total angular momentum $\overline{v_\theta} = 0$, so that $\overline{v_\theta^2} = \sigma_\theta^2$.
However, this assumption cannot be applied to the meridional direction if the system has non-zero angular momentum, leading to $\overline{v_\phi} \neq 0$.
By converting the differential operators to log scale, Equation~\ref{eq:binney} can be rewritten as
\begin{equation}\label{eq:jeans}
\sigma_{\rm r}^2 = v_{\rm c}^2(\alpha_* - 2\beta - 2\gamma_{\rm r})^{-1}.
\end{equation}

Equation~\ref{eq:jeans} indicates that the radial velocity dispersion profile $\sigma_{\rm r}(r)$ is primarily correlated with three parameters: the circular velocity profile $v_{\rm c}(r) = \sqrt{GM(<r)/r}$, the stellar density gradient $\alpha_*(r) = -d(\ln{\rho_*})/d(\ln{r})$ and the anisotropy profile $\beta(r)$.
The {\sigr} gradient $\gamma_{\rm r}(r) = d(\ln{\sigma_{\rm r}})/d(\ln{r})$ depends on the {\sigr} profile itself, and its change along the radial distance is typically small ($\pm0.3$ at maximum).
Thus, we do not consider it as one of the important parameters that describe $\sigma_{\rm r}(r)$. 
Accordingly, the role of each parameter affecting $\sigma$ profiles can be summarized as follows.
\begin{enumerate}
    \item The circular velocity, $v_{\rm c}(r)$, directly correlates with $\sigma_{\rm r}(r)$ and is a function of the enclosed mass profile, which can also be represented by the power-law slope of the {\em total} density profile $\alpha(r) = -d(\ln{\rho})/d(\ln{r})$.
    \item The {\em stellar} density power-law slope, $\alpha_*(r)$, has a negative correlation with the $\sigma_{\rm r}(r)$ profile. 
    \item The velocity anisotropy, $\beta(r)$, quantifies the dominance of either radial or tangential stellar motion and is positively correlated to the $\sigma_{\rm r}(r)$ profile. When the anisotropy increases (i.e., the velocity becomes radially biased), the velocity dispersion increases. \end{enumerate}
In Section~\ref{sec:sigm_assembly}, we observed that the slope of velocity dispersion correlates with the stellar mass and ex-situ fraction. 
Here, we demonstrate how the 3 parameters mentioned above change with stellar mass and ex-situ fraction and present their significance to the velocity dispersion profile.
\begin{figure}[t]
\includegraphics[width=0.47\textwidth]{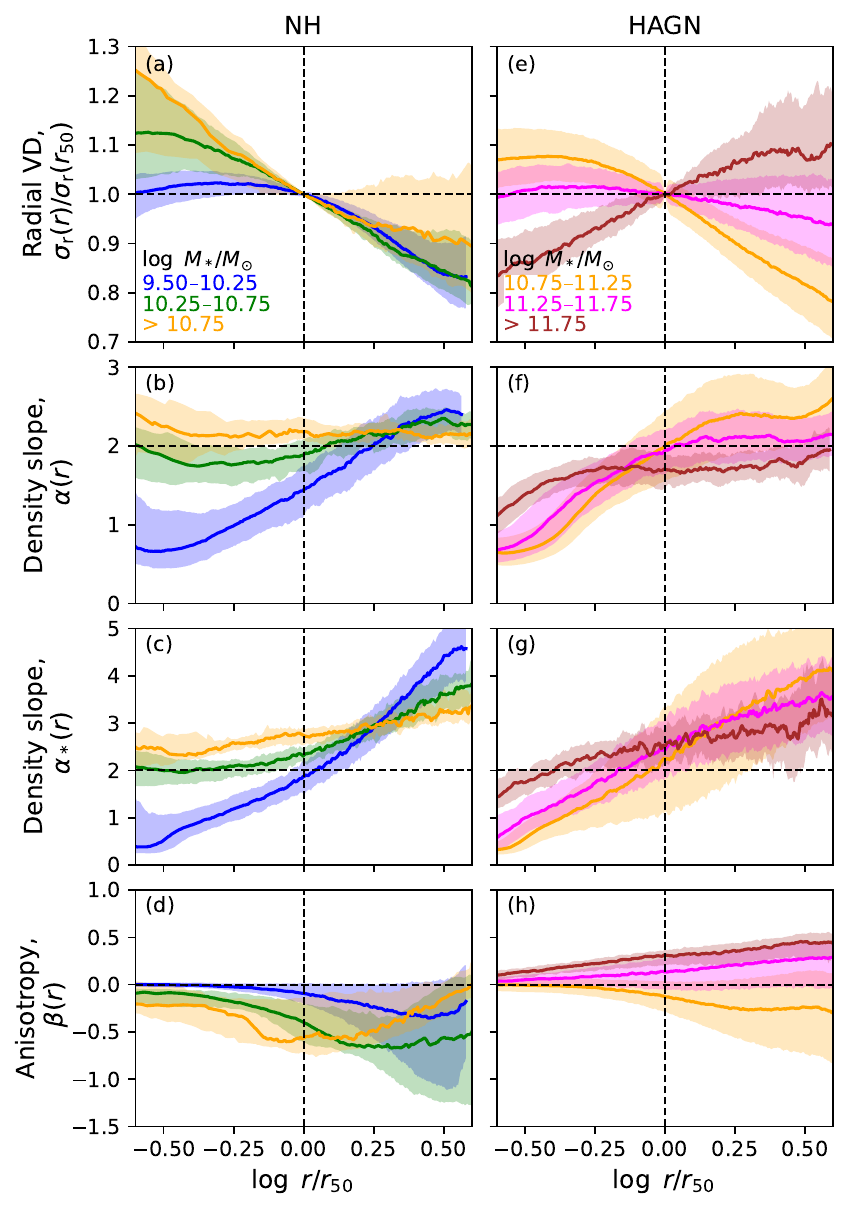}\caption{
Median radial velocity dispersion profile, normalized by its value at $r_{50}$ (1st row) and three parameters (2nd-4th row) of the Jeans' equation, namely $\alpha(r)$ (total density slope), $\alpha_*(r)$ (stellar density slope) and $\beta(r)$ (velocity anisotropy) are shown as functions of the radial distance, normalized by $r_{50}$. Each column represents NH (left) and HAGN (right) galaxies. Each line represents a distinct range of stellar mass, indicated by different colors. Shades indicate $1\sigma$ scatter.
The total and stellar density profiles are dependent on the stellar mass, which primarily drives the changes in the velocity dispersion profile.}
\label{fig:sig_mstar}
\end{figure}

Figure~\ref{fig:sig_mstar} shows the radial profiles of the {\em radial-component} velocity dispersion ({\sigr}) in the first row, and the profiles of three parameters ($\alpha$, $\alpha_*$, and $\beta$) in the second-fourth rows, as a function of radial distance.
We present the total density profile slope, $\alpha$, instead of the circular velocity, $v_{\rm c}$, for the reasons mentioned above.
Each line with a different color represents the median profile of the galaxies subsampled based on their stellar mass.
The left and right columns correspond to NH and HAGN, respectively.
In the first row, the {\sigr} profile of the NH galaxies exhibits significant mass dependence on the inner slope.
In low-mass galaxies, the inner velocity dispersion exhibits a flattened profile. 
However, as the stellar mass increases, the slope of the velocity dispersion profile transitions to a steeper, falling profile, indicating a more rapid increase in the velocity dispersion toward the inner regions of the galaxy.

In contrast, the HAGN galaxies, which cover a higher mass range ($\log(M_*/M_\sun) > 10.75$), show a clear transition of the slope with stellar mass from a negative, falling slope to a positive, rising slope.
The mass dependence of the radial {\sigr} profile (first row: panels a and e) is consistent with the trend of the projected velocity dispersion slope, as shown in Figures~\ref{fig:mstar_vs_gamma_all} and \ref{fig:mstar_vs_gamma_out}.

The second and third rows present the power-law density slopes for the total and stellar masses of the galaxies, respectively, categorized by different stellar masses.
In NH, the density profile exhibits a significant difference in the inner region.
More massive galaxies maintain a steeper central density slope.
In contrast to the NH galaxies in Panel (b), the HAGN galaxies in Panel (f) exhibit more distinct mass dependence on the outskirts.
More massive galaxies exhibit a shallower slope, indicating a more extended mass distribution on the outskirts.

The fourth row demonstrates the radial distribution of velocity anisotropy.
There is a weak dependence of the velocity anisotropy on the stellar mass in NH and HAGN.
However, there is a distinct difference in the mass dependence between NH and HAGN.
In NH, more massive galaxies have more negative values of $\beta$ and thus tangential anisotropy, indicating the presence of rotational motion.
On the contrary, the HAGN galaxies typically have positive values of $\beta$ indicative of radial anisotropy.
Overall, the change in the velocity dispersion profile with respect to the stellar mass in panels (a) and (e) appears to be primarily driven by the change in the density profile.

There appears to be some degree of consistency between NH and HAGN in the outer velocity dispersion and anisotropy profiles for common mass bins (orange lines) which correspond to the most massive galaxies of our NH sample and the least massive galaxies of our HAGN sample.
However, it is worth noting that the HAGN and NH samples exhibit different trends in the central velocity dispersion profile and density slope, even for similar mass ranges.
This can be partly attributed to the difference in the environments they represent.
HAGN, being a large-volume simulation, covers a wide variety of environments and thus includes all types of galaxies.
In contrast, NH represents a field region; thus, its massive galaxies are mostly late-type, exhibiting a dynamically hot bulge and a rotating cold star-forming disk, leading to a steeper profile and tangential anisotropy in the inner region compared to its counterpart in HAGN.
Additionally, the differences in spatial and mass resolutions between the two simulations may have led to variations in the stellar dynamics of the inner regions, which suggests that the inner kinematic information of HAGN galaxies, particularly for lower mass galaxies, may not be reliable.

\begin{figure}[t]
\begin{center}
\includegraphics[width=0.47\textwidth]{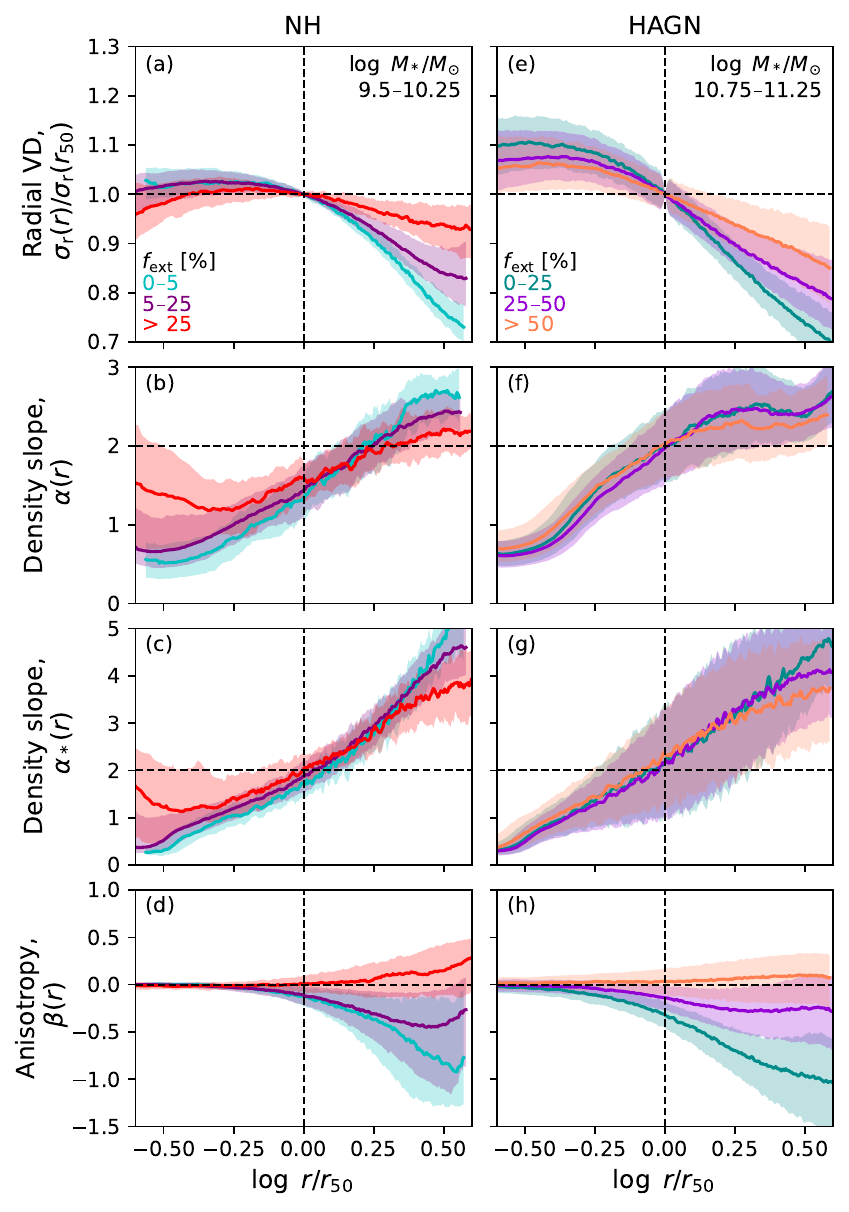}\caption{
Median radial velocity dispersion profiles and three parameters for simulated galaxies (same as Figure~\ref{fig:sig_mstar}) with different ex-situ fractions.
The variation of anisotropy profiles primarily drives the changes in the outer slopes of velocity dispersion profiles.
\label{fig:sig_fext}}
\end{center}
\end{figure}

In Figure~\ref{fig:sig_fext}, the median profiles of the parameters are plotted in each row in the same manner as in Figure~\ref{fig:sig_mstar}, except that the galaxies are now subsampled by ex-situ fraction.
To address the bias introduced by the correlation between the stellar mass and ex-situ fraction (as shown in Figure~\ref{fig:facc_slope}), we first narrow the stellar mass ranges to $\log (M_*/M_\sun) = 9.5$--10.25 for NH and $\log (M_*/M_\sun) = 10.75$--11.25 for HAGN.
These mass ranges correspond to the subsamples with the lowest masses, as shown in Figure~\ref{fig:sig_mstar}.
In the first row, the {\sigr} profiles at large radii clearly exhibit dependence on ex-situ fraction in both simulations.
In the second-third rows, no strong dependence on ex-situ fraction is observable for the total density or stellar density profiles.
In contrast, we found a strong correlation between the anisotropy profile $\beta(r)$ and ex-situ fraction (panels (d) and (h)).
In both NH and HAGN, the galaxies show a nearly isotropic velocity distribution at the inner radii, regardless of ex-situ fraction.
At large radii, high ex-situ fraction galaxies remain isotropic ($\beta(r) \sim 0$), whereas low ex-situ fraction galaxies exhibit a tangentially biased motion ($\beta(r) < 0$).
This negative slope of the velocity anisotropy profile leads to a reduction in the radial velocity dispersion ($\sigma_{\rm r}(r)$) at large radii, as shown in panels (a) and (e).
This reduction in $\sigma_{\rm r}(r)$ results in a deviation between the galaxies with different ex-situ fractions.

\begin{figure}[t]
\includegraphics[width=0.47\textwidth]{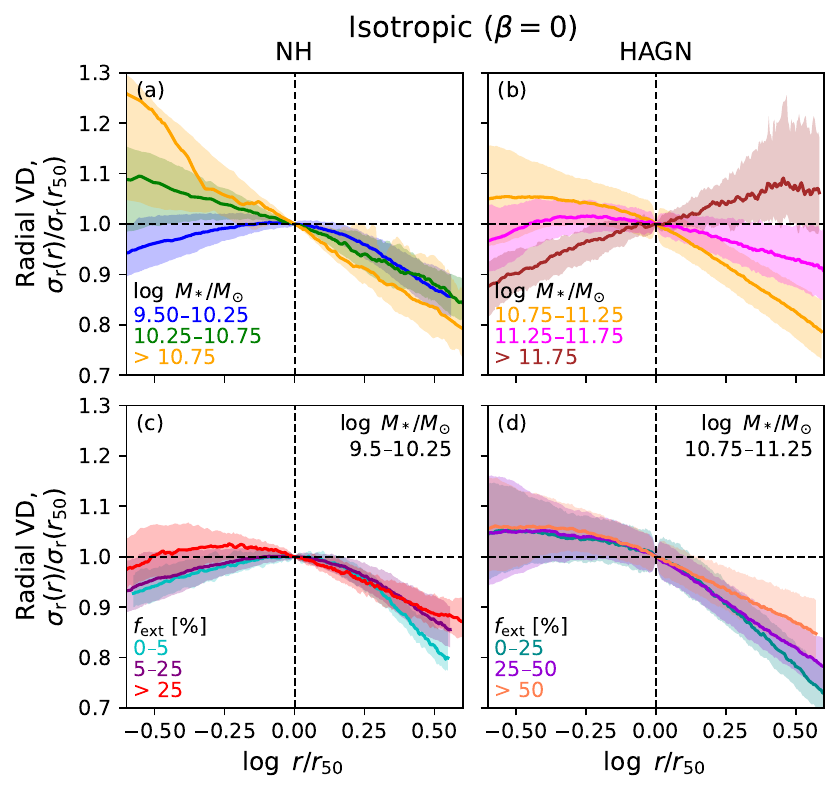}\caption{
Same {\sig} profiles as Figure~\ref{fig:sig_mstar} and Figure~\ref{fig:sig_fext}, except for assuming $\beta=0$ for all radii to eliminate effects from velocity anisotropy.
The variation of outer velocity dispersion profiles in the function of ex-situ fraction is primarily caused by velocity anisotropy.
\label{fig:sig_flatbeta}}
\end{figure}

Because these three parameters contribute to the {\sigr} profile in combination, it is not trivial to determine the direct effect of each parameter.
Therefore, to analytically remove the effect of the radial variation of the velocity anisotropy on the {\sigr} profile, we assume a flat velocity anisotropy.
More explicitly, we multiply $\sigma_{\rm r}(r)$ by $((\alpha_* - 2\beta - 2\gamma_{\rm r}) / (\alpha_* - 2\gamma_{\rm r}))^{1/2}$.
This corresponds to how the {\sigr} profile would look for the case of isotropic velocity distribution ($\beta(r)=0$).
The results are shown in Figure~\ref{fig:sig_flatbeta} binned by the stellar mass (top row) and ex-situ fraction (bottom row), respectively.
In the upper panels, a systematic variation is visible between subsamples, indicating that the stellar mass dependence still exists.
This means that the mass dependence of the {\sigr} profile cannot be explained based on the correlation between mass and velocity anisotropy alone.
In contrast, the bottom panels show that the dependence on the ex-situ fraction no longer exists after the anisotropy effect is removed.
In summary, when galaxies with different stellar masses are compared, the mass profiles (both total and stellar) are the key factors driving the change in the velocity dispersion profile.
However, when the stellar mass is fixed, a radial variation in the velocity anisotropy, which is closely related to ex-situ fraction, is the key driver that causes a change in the outer {\sigr} profile gradient.

\section{Discussion}
\subsection{Impact of ex-situ fraction and radial anisotropy}\label{sec:discuss}

\begin{figure}[t]
\begin{center}
\includegraphics[width=0.47\textwidth]{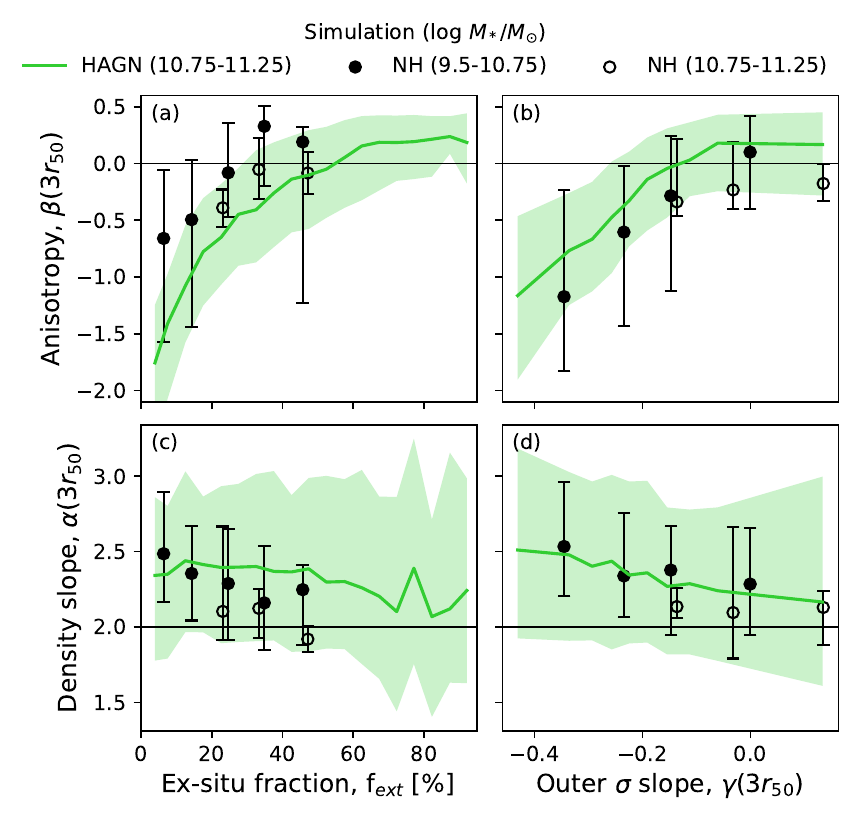}
\caption{Velocity anisotropy (top panels) and total mass density slope (bottom panels) at $3\,r_{50}$ as a function of ex-situ fraction (left panels) and outer velocity dispersion slope (right panels), drawn for HAGN galaxies (green lines) and NH galaxies (black markers). Shaded region and error bars indicate $1\sigma$ scatter.
Velocity anisotropy shows a good correlation with ex-situ fraction and outer $\sigma$ slope.
\label{fig:facc_beta_alpha}}
\end{center}
\end{figure}

Our results indicate that there is a wide variation in the outer velocity dispersion slopes among both simulated and observed galaxies. 
This diversity is primarily caused by the different fractions of ex-situ stars present in the galaxies (see Figure~\ref{fig:facc_slope}).
Furthermore, galaxies with higher ex-situ fractions exhibit less tangentially biased velocity distribution (see Figure~\ref{fig:sig_fext}-d \& h).
In Figure~\ref{fig:facc_beta_alpha} we further investigate the roles of $\alpha$ and $\beta$.
We take the HAGN galaxies as our main sample here because they cover a wide range of galaxy types and environments.
In order to minimize the mass effect while securing a sufficiently large sample for statistics, we limit the HAGN galaxy sample to $\log M_*/M_\sun = 10.75$--11.25.
For reference, we show the NH sample, too.
We divide the NH sample by mass: the open circles show the same mass range as that used for the HAGN galaxies in this diagram, and the closed circles show the lower-mass sample which represents the majority of NH. 
The figure shows anisotropy and density slope measured at $3\,r_{50}$ of simulated galaxies with respect to ex-situ fraction and {\gout}.
We find a strong correlation between the anisotropy and ex-situ fraction (panel (a)), while the correlation between the density slope and ex-situ fraction is weak at best (panel (c)).
Similar trends are observed when compared with the outer slope of the velocity dispersion profile (panels (b) and (d)).
Panel (c) shows that ex-situ fraction does not affect the outer density slope, which has already been demonstrated in Figure~\ref{fig:sig_fext} (second-third rows).
A change in {\gout} occurs primarily through ex-situ stars via a change in anisotropy rather than a change in density slope.

In the case of most massive galaxies, it becomes difficult to investigate the effects of accretion because they exhibit little variation in ex-situ fraction among themselves.
For example, the HAGN galaxies of $\log (M_*/M_\sun) \gtrsim 11.6$ have an ex-situ fraction of 80--100\% (see Figure~\ref{fig:facc_slope}).
Their outer velocity dispersion profiles are affected both by density slopes (i.e., mass) and anisotropy (see Figure~\ref{fig:sig_mstar}).
As a result, they exhibit 1) higher slopes in outer velocity dispersion profiles compared to lower mass counterparts, accompanied by 2) radial anisotropy.
This finding is consistent with observations of massive early-type galaxies, which also display higher slopes of velocity dispersion and positive values of the $h_4$ parameter \citep{2018MNRAS.473.5446V, 2018MNRAS.477..335L}.

However, it is widely accepted that the radial anisotropy causes the projected velocity dispersion to have a falling profile \citep{1993MNRAS.265..213G}, which appears to contradict our finding of higher slopes in outer velocity dispersion profiles mentioned above.
For example, \cite{2020MNRAS.496.1857L} demonstrated an expected anti-correlation between the anisotropy parameter $\beta$ and the velocity dispersion slope.
However, they assumed a constant anisotropy, which differs from the radially varying anisotropy observed in our simulated galaxies.

In fact, given that all our simulated galaxies have a more isotropic velocity distribution ($\beta=0$) toward the center, the effect of radial anisotropy gradient is the opposite of that expected from the projection effect.
According to Equation~\ref{eq:jeans}, with the total and stellar mass profile fixed, the velocity dispersion should move towards a rising profile if radial anisotropy increases with increasing radii (Figure~\ref{fig:sig_mstar}).
This can be interpreted in a way that galaxies with high ex-situ fractions show an intrinsically (i.e., in 3-D) rising velocity dispersion profile that is significant enough to overwhelm the projection effect.
It is natural to expect that the accreted stars will maintain momentum along the infalling path, making them follow a radially biased orbit, and also mainly reside in the outskirt due to the high kinetic energy gained from the potential of the accreting galaxy.

\begin{figure}[t]
\begin{center}
\includegraphics[width=0.47\textwidth]{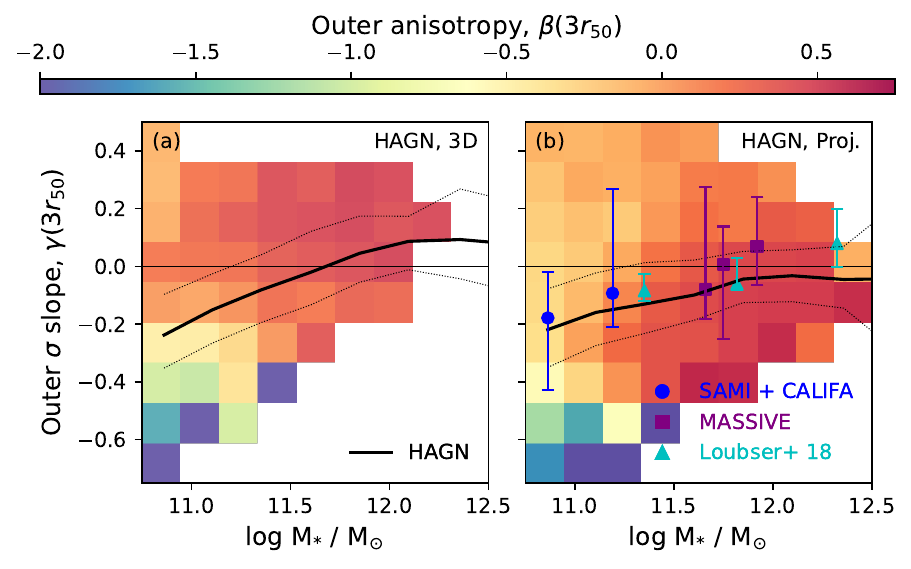}
\caption{Same as Figure~\ref{fig:facc_slope}, except for the background color pixels indicate anisotropy parameter measured at $3\,r_{50}$.
More massive galaxies exhibit more radial velocity anisotropy and higher velocity dispersion slopes, which are consistent with observations.
\label{fig:beta_slope}}
\end{center}
\end{figure}

Figure~\ref{fig:beta_slope} displays the same plot as Figure~\ref{fig:facc_slope} but with the background color replaced with the anisotropy parameter at $3\,r_{50}$.
The intrinsic slope generally shows a vertical gradient of increasing $\beta$ with the outer slope.
This can be understood as a competitive contribution between the in-situ stars and the ex-situ stars, with the former having tangential anisotropy due to disk-like kinematics and the latter having radial anisotropy due to the infalling velocity during the accretion or merger.
When the projection effect is taken into account, the outer slope of the most massive galaxies shifts toward a falling profile. The upper part of the figure (rising and flat slope) exhibits a weak vertical gradient with different directions to that of ex-situ fraction.
This is primarily driven by a collective, but differential shift of galaxies toward lower outer slopes depending on their degree of radial anisotropy.

\subsection{Comparison with other literature}
Our result is qualitatively consistent with \cite{2022MNRAS.513.6134W}, which employed a similar technique to ours in a cosmological simulation IllustrisTNG \citep{2018MNRAS.473.4077P} based on moving mesh hydrodynamics.
They focused on ETGs for comparison with counterparts of the MASSIVE dataset \citep{2014ApJ...795..158M}.
They found that the outer velocity dispersion slope, despite the difference in the definition of the radius where the slope is measured ($1.5r_{\rm eff}$) compared to ours ($3\,r_{50}$), is positively correlated with stellar mass, velocity anisotropy, and major merger fraction, which is in line with the ex-situ fraction.
However, on the contradicting observation of positive $\sigma$ gradient with positive $h_4$ parameter, they suggest $h_4$ is uncorrelated to the stellar orbital anisotropy.
This differs from our explanations based on the variation of orbital anisotropy.
This can be a subject of a future study.

The study of \cite{2023MNRAS.520.5651C} also utilizes the IllustrisTNG simulation and compares its velocity dispersion profile with the MaNGA IFU survey \citep{2015ApJ...798....7B,2016AJ....152..197Y}, providing velocity dispersion profiles of in-situ and ex-situ stars in galaxies.
Their result indicates that the overall velocity dispersion of ex-situ stars shows a slightly higher value than in-situ stars (Figure 3, 4th row in their result), which is consistent with our result (Figure~\ref{fig:sig_orig} and \ref{fig:sig_orig_hagn}.
However, for larger radii and more massive galaxies, the difference appears not to be as significant as ours.
Further investigation is needed to explore this aspect in more detail.

\subsection{Caveats}
There are several caveats in this study.
Beam-smearing is one of the major obstacles in measuring velocity dispersion profiles in observations.
In our main results, we did not account for the beam-smearing effect on 2-D velocity dispersion profiles for simulated galaxies.
This could potentially introduce bias to the measurement of slope in the velocity dispersion profile when compared with observations.
Another related issue is the variation in beam size applied to different observational datasets, leading to potential inconsistencies in comparisons.
We discuss in Appendix \ref{sec:beam} the impact of the artificial beam-smearing effect applied to simulated galaxies.
The result suggests that the impact of beam-smearing is not large enough to alter the main trends found in Figures~\ref{fig:mstar_vs_gamma_all} and \ref{fig:mstar_vs_gamma_out}.
From the result, we also anticipate that the influence of differences in beam sizes between observations will be small.

Dust absorption is not considered in estimating the luminosity-weighted velocity dispersion profile from simulated galaxies.
This could result in an unusually bright central region, reducing the effective radius of galaxies, as well as affecting the central velocity dispersion profile.
This problem is particularly prominent for the high-mass NH galaxies, which have both high resolution and bright central bulges.

The use of Voronoi cells may result in an overestimation of velocity dispersion in sparse regions, particularly where cell size increases and the spatial gradient of the stellar group velocity becomes significant.
This issue is prevalent in the measurement of 2-D velocity dispersion profile in both observation and simulation.
While it may be particularly influential for the outskirts of low-mass galaxies flattening the velocity dispersion profiles, the actual bias may not be very significant considering the large scatter in their outer $\sigma$ slope, as shown in Figure~\ref{fig:mstar_vs_gamma_out}.

Throughout the paper, we consistently employed a fixed shape and position angle for the ellipses of each galaxy in the 2-D velocity dispersion measurement, both for observed and simulated galaxies.
The purpose of employing ellipses is to account for the effect of inclination on the binning of stellar information, ensuring that the binning optimally reconstructs the 3-D shells surrounding the galaxy.
This is based on the assumption that visually elongated galaxies possess an oblate shape or are composed of ideal thin disks.
It should be acknowledged that some galaxies have triaxial shapes \citep{1978ApJ...222....1K,2007ApJ...670.1048K, 2022ApJ...930..153S} or warped disks \citep{1976A&A....53..159S,1990ApJ...352...15B} that likely have a radial variation in shape and position angle when fitted by ellipses.
We do not account for this effect, and as a result, our velocity dispersion measurements are subject to uncertainties related to it.

Throughout the kinematic analysis of the 3-D velocity dispersion profile, we assumed each simulated galaxy to have a spherically symmetrical mass distribution and to be in an equilibrium state where inflow and outflow are equal.
This simplification may not fully represent the actual structure of galaxies.
However, we observed that Jeans' equation reasonably well reproduces the expected velocity dispersion profile of simulated galaxies from a given density and velocity anisotropy profile, indicating that our assumptions are valid to some extent.

\section{Conclusion}\label{conclusion}

In this study, we have used two cosmological simulations, Horizon-AGN (for $\log(M_*/M_{\sun}) > 10.75$ at $z=0.017$) and NewHorizon (for $\log(M_*/M_{\sun}) > 9.5$ at $z\lesssim1$), to investigate the velocity dispersion profiles of galaxies.
We have measured the slopes of the velocity dispersion in the 2-D plane at the inner ($0.5\,r_{50}$) and outer ($3\,r_{50}$) radius and compared them with integral field spectroscopy data from two surveys, SAMI (for $z<0.115$) and CALIFA (for $z<0.03$) , for galaxies within the range of $\log(M_*/M_{\sun}) > 9.5$.
Using the versatility provided by numerical simulations, we have investigated the variations in the velocity dispersion profiles in connection with the spatial age distribution, mass profile, and velocity anisotropy, and found their close connection to the stellar mass and ex-situ fraction of galaxies.
Our main results can be summarized as follows.

\begin{enumerate}
    \item The inner slope of the velocity dispersion profile shows a ``V-shaped" trend with stellar mass in both simulated and observed samples.
    The slope becomes more negative (moving towards a falling profile) with increasing stellar mass until $\log(M_*/M_\sun) \sim 11$, and appears to bounce back thereafter (Figure~\ref{fig:mstar_vs_gamma_all}).
    The outer slopes of velocity dispersion profiles show a large variation.
    Lower-mass galaxies ($\log(M_*/M_\sun)\lesssim11$) exhibit no strong dependence on mass.
    There is a hint of an increasing slope with stellar mass in more massive galaxies ($\log(M_*/M_\sun)\gtrsim11$), where the observational and simulation data are in agreement with each other (Figure~\ref{fig:mstar_vs_gamma_out}).
    \item The shape of the inner velocity dispersion profile of a galaxy is determined by the relative radial distribution of young and old stellar populations. The change in the inner slope of the velocity dispersion profile with increasing stellar mass is driven by a transition in the order of the star formation process, from outside-in to inside-out (Figure~\ref{fig:sig_orig}).
    \item Ex-situ stars exhibit higher velocity dispersion than in-situ stars and are more predominant in the outer region of the galaxy, hence increasing the mean velocity dispersion at the galactic outskirts (Figure~\ref{fig:sig_orig}). As a result, galaxies with higher ex-situ fractions have higher outer velocity dispersion slopes (Figure~\ref{fig:facc_slope}).
    \item The stellar mass dependence of the velocity dispersion profile is primarily driven by changes in the mass profile (Figure~\ref{fig:sig_mstar}). However, the diverse outer velocity dispersion slope among galaxies with similar masses is primarily driven by the radial variation of anisotropy (Figure~\ref{fig:sig_fext}), which tightly correlates with ex-situ fraction (Figure~\ref{fig:facc_beta_alpha}). On the other hand, ex-situ fraction does not affect the outer density profile (Figure~\ref{fig:facc_beta_alpha}).
    \item We found that most massive ($\log(M_*/M_\sun)\gtrsim11.6$) galaxies exhibit a very high ex-situ fraction (80--100\%, Figure~\ref{fig:facc_slope}), resulting in a higher outer velocity dispersion slope and radial anisotropy, which is consistent with observations.
\end{enumerate}

This investigation suggests that the recent star formation and past assembly history of a galaxy contribute to shaping its velocity dispersion profile.
However, this perspective may be viewed as an oversimplification, given the various complications involved, such as environmental effects, the presence of supermassive black holes, and different modes of star formation.
Pinning down the exact processes observationally proves challenging, but exploring them in numerical simulations could offer valuable insights. This topic holds promise for future studies.

The result presented in this paper is largely independent of redshift.
This is indicated in Figures~\ref{fig:mstar_vs_gamma_all} and \ref{fig:mstar_vs_gamma_out}, by comparing the final snapshot ($z=0.17$) and the full sample ($z<1$) of NH galaxies.
This is because, while redshift evolution may alter the absolute values of the velocity dispersion of galaxies \citep{2020MNRAS.498.1101C}, our analysis focuses on measuring the log-log slope of $\sigma(r)$ at inner and outer radii.
The trend we find is more dependent on the specific history of individual galaxies rather than on the global evolution of galaxies with cosmic time.
In addition, introducing an artificial beam-smearing effect into the simulation by applying a Gaussian kernel-based measurement of the velocity dispersion profile does not significantly change our conclusion, as indicated in Appendix \ref{sec:beam}.

We propose two parameters that can potentially serve as indicators for measuring the ex-situ fraction of galaxies: the outer slope of the velocity dispersion profile as a weak indicator, and the radial anisotropy of the velocity distribution as a strong one.
The former can be measured relatively easily from stellar absorption line profiles, although it is still challenging to get it done on outer regions of galaxies.
Measuring the latter presents an even greater challenge.
Measuring the higher order Gauss-Hermite moment $h_4$ requires even more accurate spectroscopy. 
Future observations should aim to perform these measurements for a greater sample of galaxies, covering wider ranges of mass and environments, than what is available now.

Analyzing the kinematic properties of today's galaxies and inferring past events are pivotal to understanding the role of mergers and accretions in the evolution of galaxies.
The measurement of ex-situ fraction for example links the kinematic properties of galaxies at the galactic scale and their hierarchical assembly histories in a large-scale environment.

One interesting feature of this work is the presence of an upturn in the trend of inner velocity dispersion slopes with stellar mass (Figure~\ref{fig:mstar_vs_gamma_all}).
Both observations and simulations consistently suggest negative correlations for lower masses and positive correlations for higher masses.
This strongly indicates the presence of a turning point around $\log(M_*/M_\sun)\sim11$.
The ``V-shaped" trend is likely the result of two effects impacting the inner slope in different directions.
Lower-mass galaxies, with low ex-situ fractions, are more influenced by the properties of in-situ stars. The differential distribution of young and old in-situ stars results in negative correlations of the inner slope with mass. However, above a certain stellar mass, ex-situ stars that are more abundant in more massive galaxies may infiltrate the inner regions of galaxies, influencing the inner velocity dispersion profiles,  producing the upturn in the trend.

It is also intriguing that the stellar mass of the upturn in the trend corresponds to the point where the galaxy transitions from a star-formation-dominated phase to a merger-dominated phase in the size evolution \citep{2015ApJ...813...23V}
This is consistent with observations showing that the upturn is not present for late-type galaxies, which are not star-formation quenched yet and have not undergone dry mergers
(Figure~\ref{fig:mstar_vs_gamma_ltg}).
For early-type galaxies, the turning point roughly corresponds to the stellar mass where the distinction between fast and slow rotators occurs \citep{2007MNRAS.379..401E,2011MNRAS.414..888E}.
This is consistent with our scenario of in-situ to ex-situ transition, as slow rotators are expected to be a result of dry mergers \citep{2017ApJ...837...68C}.

The use of two simulations (NH and HAGN) in combination in this study reproduced the observed upturn in the inner slope.
However, the significant difference in astrophysical prescriptions and (spatial and mass) resolutions between the two simulations poses a major caveat, making it challenging to interpret the results and determine. 
This emphasizes the need for high-resolution simulations that are capable of reproducing a diverse galaxy population covering a wide range of stellar masses and environments under a single astrophysical prescription.
If the current computing resources are limited to perform such a simulation in one piece, it would be a good strategy to run separate zoom-in simulations using identical physical ingredients and resolutions.

\section*{Acknowledgements}
% NH
We thank the anonymous referee for the comments that clarified the manuscript greatly. This work was granted access to the HPC resources of CINES under the allocations  c2016047637, A0020407637 and A0070402192 by Genci, KSC-2017-G2-0003 by KISTI, and as a “Grand Challenge” project granted by GENCI on the AMD Rome extension of the Joliot Curie supercomputer at TGCC.
The large data transfer was supported by KREONET which is managed and operated by KISTI. 
% SKY
S.K.Y. acknowledges support from the Korean National Research Foundation (NRF-2020R1A2C3003769). 
% SO
S.O. acknowledges support from the NRF grant funded by the Korean government (MSIT) (RS-2023-00214057). 
Parts of this research were conducted by the Australian Research Council Centre of Excellence for All Sky Astrophysics in 3 Dimensions (ASTRO 3D), through project number CE170100013.
% TK
T.K. was supported in part by the NRF (2020R1C1C1007079).
% MP
% CP
This work is partially supported by the grant Segal ANR-19-CE31-0017 of the French Agence Nationale de la Recherche and by the National Science Foundation under Grant No. NSF PHY-1748958. 
% general
This study was also funded by the NRF-2022R1A6A1A03053472 grant and the BK21Plus program.
S.H. and S.K.Y. acted as the corresponding authors.

%\end{acknowledgements}

%\nocite{*}
\bibliography{ms}{}
\bibliographystyle{aasjournal}

\appendix
\restartappendixnumbering
\section{Assessing the quality of broken power-law fit}\label{sec:fitting_quality}
\begin{figure*}[hbt]
\centering
\includegraphics[width=0.95\textwidth]{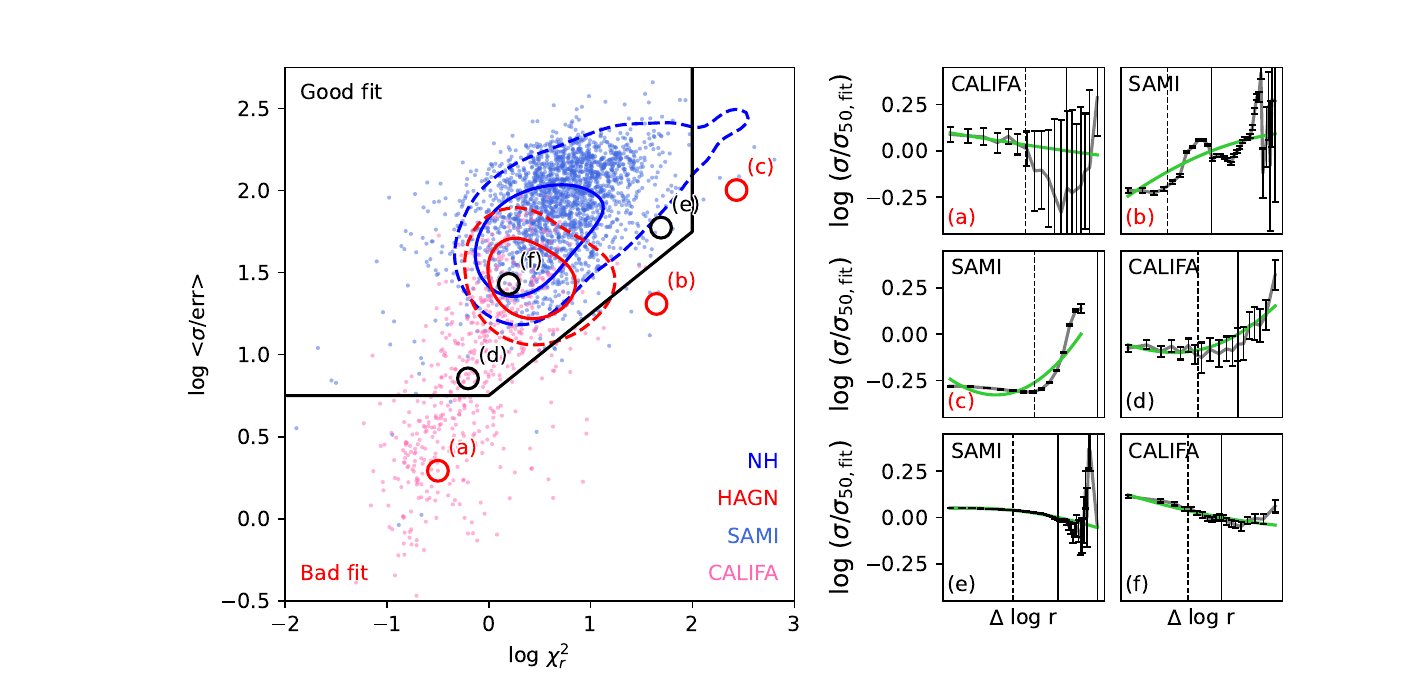}
\caption{
Distribution of reduced chi-square value of the fitting (x-axis), compared to mean $\sigma$ to error ratio (y-axis). Pink and blue points indicate power-law fit from SAMI and CALIFA, respectively.
Red and blue contours indicate simulated galaxies from HAGN and NH, respectively, with 1-sigma (solid lines) and 2-sigma (dashed lines) confidence levels of the distribution.
A black solid line indicates the threshold to select samples with reliable fits.
In the left panels, 3 examples of bad fits ((a), (b), and (c)), and good fits ((d), (e), and (f)) are presented, indicated with red and black markers, respectively, on the corresponding position of the left panel.
The fitted broken power law curve (green) is compared to data profiles (gray) and errors (black).
Vertical lines indicate $r_{50}$ (solid) and $0.5\,r_{50}$ (dashed) of the galaxy.
\label{fig:fitting_quality}}
\end{figure*}

This section describes how we select the galaxy profiles for our analysis from the SAMI and CALIFA databases based on the goodness of fit.
In Figure~\ref{fig:fitting_quality}, the distribution of fitting results is shown with reduced chi-square (x-axis) and mean $\sigma$ to error ratio (y-axis) in the left panel, and six examples of broken power law fits are shown in the right panels.
Panels corresponding to red markers ((a), (b), and (c)) represent examples of bad fit, while panels with black markers ((d), (e), and (f)) indicate good fits.
We used the following criteria to select good fits:
\begin{equation}\label{eq:criterion}
\log{\langle\sigma / \rm{err}\rangle} > 0.75 \ \  \text{and} \ \ 2\log{\langle\sigma / \rm{err}\rangle}-\log{\chi_r^2} > 1.5 \ \  \text{and} \ \ \log{\chi_r^2} < 2,
\end{equation}
considering the $\sigma$-to-error ratio and reduced chi-squared.
The first part of the criteria, $\log{\langle\sigma / \rm{err}\rangle} > 0.75$ is applied to rule out data with too large error on the measurement of the velocity dispersion profile.
The panel (a) shows an example of this case.
Although the fit has a small $\chi_r^2$, the large error causes the broken power-law fit to deviate significantly from the data points.
The second part of the criteria implies the ruling out of common bad fits that has both low quality of the data and low $\chi_r^2$, which is shown in panel (b) as an example.
The third part of the criteria, $\log{\chi_r^2} < 2$, sets a limit for the quality of the fit.
Even though the data has a small error ($\langle\sigma / \rm{err}\rangle$), a significant deviation in the shape of the profile from the broken power-law profile results in a very high $\chi_r^2$ and is considered a bad fit.
The three other panels, (d), (e), and (f) represent examples of good fits that satisfy the selection criteria.
It is worth noting that panel (e) and (f) include several data points that seem to deviate from the fitting curve.
However, this apparent inconsistency is attributed to the fitting algorithm, which disregards bins with relatively large errors.
The same criterion is applied to simulated galaxies (NH, HAGN), which are represented as contours, and predominantly satisfy the criterion.

\restartappendixnumbering
\section{Inner velocity dispersion slope of galaxies with different morphology}\label{sec:morph}

\begin{figure*}[t]
\centering
\includegraphics[width=0.95\textwidth]{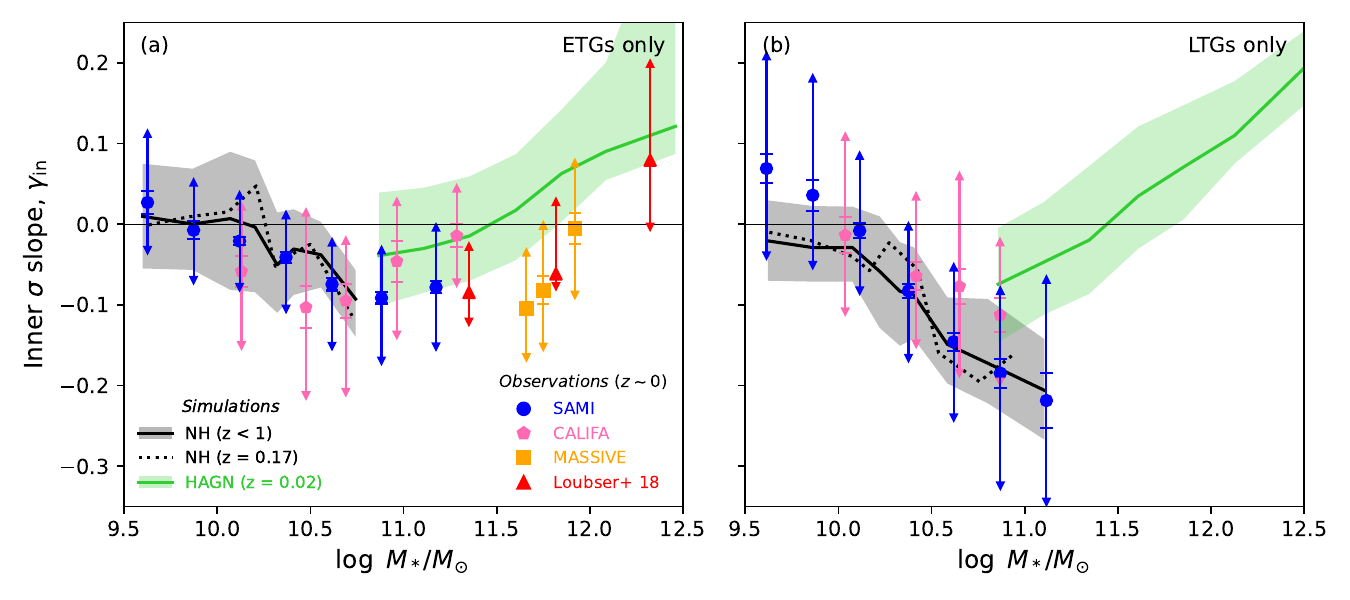}
\caption{
The inner velocity dispersion slope of ETGs and LTGs measured at $0.5\,r_{50}$ as a function of galaxies' stellar mass in the same format as Figure~\ref{fig:mstar_vs_gamma_all} in the main text. Lines in each panel show simulated galaxies in HAGN and NH classified as ETG and LTGs based on the criterion $(v/\sigma)_{\rm 3D} = 1$ (see the text for more details), with shades indicating the $1\sigma$ scatter. The NH galaxies are in good agreement with the trend of late-type galaxies in observation. To ensure the consistency of kinematical classification, we excluded two galaxies with counter-rotating features from the NH sample.
\label{fig:mstar_vs_gamma_ltg}}
\end{figure*}

Figure~\ref{fig:mstar_vs_gamma_ltg} shows for reference the inner profiles of velocity dispersion of the simulated galaxies in the same format as Figure~\ref{fig:mstar_vs_gamma_all} in the main text but for observed early- and late-type galaxies (ETGs and LTGs) separated.
For SAMI and CALIFA, galaxy morphologies are determined through visual classification.
We employed the morphology information from \cite{2016MNRAS.463..170C, 2021MNRAS.505..991C} for SAMI and \cite{2019ApJ...880..149P} for CALIFA.
We considered ellipticals (E) and leticulars (S0) as ETGs, and the rest as LTGs.
We use the criterion $(v/\sigma)_{\rm 3D}=1$ to distinguish ETGs and LTGs (or dispersion- and rotation-dominated galaxies) following Dubois et al. (2016, MNRAS, 463, 3948), where $(v/\sigma)_{\rm 3D}$ indicates the ratio between maximum stellar rotation speed and 3-dimensional stellar velocity dispersion.
We note that the criterion is for the rough classification of simulated galaxies and may not be indicative of visual morphology in some cases.
For example, a large fraction of HAGN late-type galaxies in Panel (b) may actually be fast-rotating elliptical or lenticular galaxies.
The massive NH galaxies are mostly late-type in many aspects because of being in the field environment.
The most massive galaxies in the HAGN simulations (green lines and shades) show similar trends regardless of the morphological classification used.

\restartappendixnumbering
\section{Slope of velocity dispersion with different PSF size}\label{sec:beam}
\begin{figure*}[t]
\centering
\includegraphics[width=0.47\textwidth]{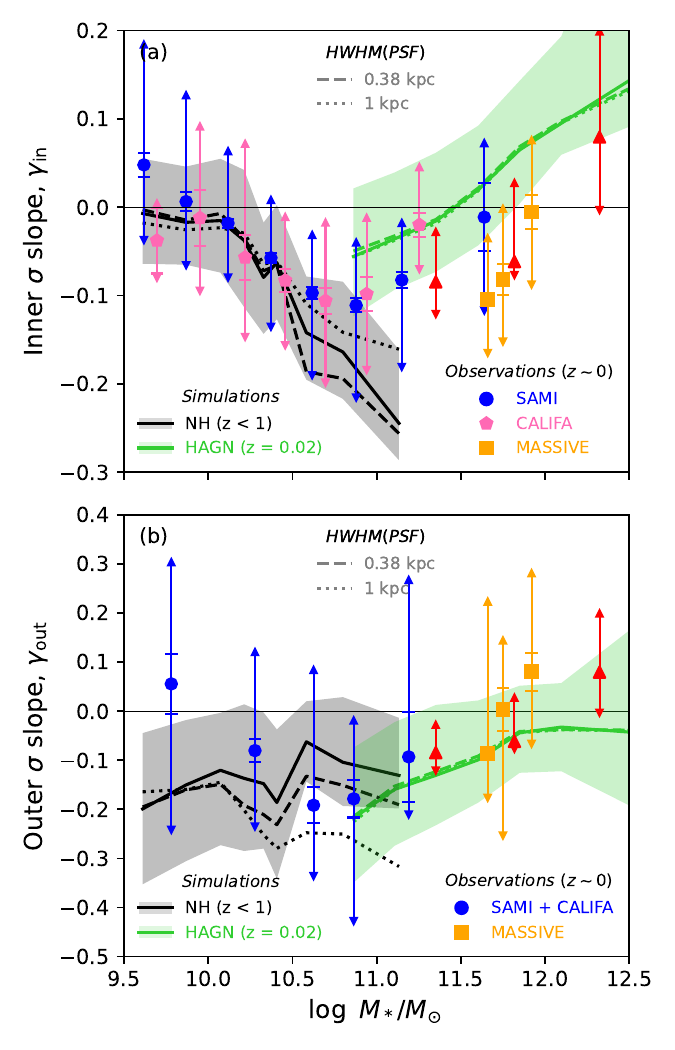}
\caption{
Similar to Figures~\ref{fig:mstar_vs_gamma_all} and \ref{fig:mstar_vs_gamma_out}, the simulated data includes the results of Gaussian convolution with two different kernel sizes on the 2-D plane, applied to the line-of-sight velocity distribution.
HWHM of 0.38 (dashed lines) and 1{\kpc} (dotted lines) correspond to the physical PSF sizes for CALIFA and SAMI galaxies, respectively, measured based on the mean redshift and angular resolution of the data.
The solid lines represent the original slopes without the beam-smearing effect.
\label{fig:mstar_vs_gamma_bs}}
\end{figure*}
To investigate the effect of beam-smearing, we present Figure~\ref{fig:mstar_vs_gamma_bs}, which shows the inner and outer slope of the velocity dispersion (same as Figure~\ref{fig:mstar_vs_gamma_all} and \ref{fig:mstar_vs_gamma_out}).
The solid lines represent the original data from previous figures.
The dashed and dotted lines represent simulated data with Gaussian smoothing applied, to consider the effect of two sizes of the point spread function (PSF) which correspond to the half-width at half-maximum (HWHM) of CALIFA (0.38{\kpc}) and SAMI (1{\kpc}), respectively.
The PSF-applied data shows differences only at the high-mass end of NH galaxies.
The inner and outer slopes are affected by $\sim25\%$ and $\sim50\%$, respectively, at $\log(M_*/M_\sun)\sim11$ of NH, where the effect is most pronounced.
This is likely to be a consequence of the changes to the velocity dispersion profile being particularly influential for the galaxies with a strong central peak in the velocity dispersion profile.
Because the luminosity of the bright central parts of massive NH galaxies is dispersed to surrounding regions, the effect may also have been amplified, thereby influencing the luminosity-weighted chi-square fitting of the broken power-law profile.
\end{document}